\documentclass[aps,pre,footinbib,showpacs,twocolumn]{revtex4-1}

\usepackage{amssymb,latexsym,mathrsfs}
\usepackage{amsfonts}
\usepackage{graphicx}
\usepackage{lmodern}
\usepackage{fix-cm}
\usepackage{xfrac}
\usepackage{color}    
\usepackage{textcomp}
\usepackage{amsmath}
\usepackage{subfigure}
\usepackage{float}
\usepackage{datetime}
\usepackage{hyperref}
\hypersetup{
    colorlinks=true,
    linkcolor=red,
    filecolor=magenta,      
    urlcolor=cyan,
}
\usepackage{tabularx,ragged2e,booktabs}

\newcommand{\be}{\begin{equation}}
\newcommand{\ee}{\end{equation}}
\newcommand{\bearr}{\begin{array}}
\newcommand{\enarr}{\end{array}}

\def\bea{\begin{eqnarray}}
\def\eea{\end{eqnarray}}
\def\ba{\begin{array}}
\def\ea{\end{array}}

\begin{document}

\title{Spatio-temporal spread of perturbations in a driven dissipative Duffing chain: an OTOC approach}
 
\author{Amit Kumar Chatterjee}\email{amit.chatterjee@icts.res.in}
\affiliation{International Centre for Theoretical Sciences, Tata Institute of Fundamental Research, Bengaluru -- 560089, India}

\author{Anupam Kundu}\email{anupam.kundu@icts.res.in}
\affiliation{International Centre for Theoretical Sciences, Tata Institute of Fundamental Research, Bengaluru -- 560089, India}

\author{Manas Kulkarni}\email{manas.kulkarni@icts.res.in}
\affiliation{International Centre for Theoretical Sciences, Tata Institute of Fundamental Research, Bengaluru -- 560089, India}


\date{\today}
\begin{abstract}
Out-of-time-ordered correlators (OTOC) have been extensively used as a major tool for exploring quantum chaos and also recently, there has been a  classical analogue. Studies have been limited to closed systems. 
In this work, we probe an open classical many-body system, more specifically, a spatially extended driven dissipative chain of coupled Duffing oscillators using the classical OTOC to investigate the spread and growth (decay) of an initially localized perturbation in the chain. Correspondingly, we find three distinct types of dynamical behavior, namely the {\it sustained chaos, transient chaos} and {\it non-chaotic region}, as clearly exhibited by different geometrical shapes in the heat map of OTOC. 
To quantify such differences, we look at  {\it instantaneous speed} (IS),  {\it finite time Lyapunov exponents} (FTLE) and {\it velocity dependent Lyapunov exponents} (VDLE) extracted from OTOC.  Introduction of these quantities turn out to be instrumental in diagnosing and demarcating different regimes of dynamical behavior. To gain control over open nonlinear systems, it is important to look at the variation of these quantities with respect to parameters.  As we tune drive, dissipation and coupling, FTLE and IS exhibit  transition between sustained chaos and non-chaotic regimes with intermediate transient chaos regimes and highly intermittent sustained chaos points. In the limit of zero nonlinearity, we present exact analytical results for the driven dissipative harmonic system and we find that our analytical results can very well describe the non-chaotic regime as well as the late time behavior in the transient regime of the Duffing chain.  We believe, this analysis is an important step forward towards understanding nonlinear dynamics, chaos and spatio-temporal spread of perturbations in many-particle open systems.
\end{abstract}

\maketitle


\section{Introduction}
\label{intro}

Chaotic and regular motion and transition between them with variation of tunable parameters has always been a central issue of interest in the context of dynamical systems. The fact that immense sensitivity to arbitrarily small perturbations in initial conditions and system parameters may result in complex dynamical behavior has led to extensive studies of chaos in numerous classical \cite{Lorenz_1993,Strogatz_1994} as well as quantum model systems \cite{Gutzwiller_1990,Haake_1991}. Needless to mention that chaos,  being ubiquitous, has found applications in various fields starting from atmospheric sciences \cite{Lorenz_1993,Zeng_1993,Huppert_1998,Selvam_2010}, chemical sciences  \cite{Field_1993,Gaspard_1998,Gaspard_1999,Srivastava_2013}, biological sciences \cite{Skinner_1994,Lloyd_1995,Ditto_1996,Lesne_2006} and technological electro-mechanical devices   \cite{Tchana_2008,Kwuimy_2008,Bao_2011,Chau_2011}.

Chaotic behavior in classical systems is diagnosed with the aid of Lyapunov exponent (LE), $\lambda$ which characterizes the rate of separation of initially infinitesimally close trajectories at large times. Depending on the sign of $\lambda$, the dynamics is classified as chaotic ($\lambda >0$) and non-chaotic/regular ($\lambda \leq 0$).
In addition to this, the phenomenon of chaos is examined using  
concepts such as phase-space portraits, Poincare sections, Bifurcation diagrams, power spectrum analysis to name a few \cite{Strogatz_1994,Gutzwiller_1990}. 
Most of the work along this line has been restricted to systems involving single \cite{Wei_1997} or very few degrees of freedom at best \cite{Kozlowski_1995}. 

In case of extended systems involving many degrees of freedom, there have been interesting studies concerning not only growth of small localized initial separation but also their spread in space. 
Examples include, propagation of chaos in reaction-diffusion systems \cite{Vastano_1988, Wacker_1995}, coupled-map lattices \cite{Lepri_1996,Lepri_1997}, Fermi-Pasta-Ulam (FPU) chain \cite{Giacomelli_2000,Pazo_2016}, complex Ginzburg-Landau system and the Gray-Scott network \cite{Stahlke_2011}, where both Lyapunov exponents and spatial propagation of perturbation are discussed 
in the contexts of computing 
time delayed mutual information and redundancy \cite{Vastano_1988}, defining both temporal as well as spatial Lyapunov exponents \cite{Lepri_1996}, introducing entropy potential \cite{Lepri_1997}, convective Lyapunov spectrum \cite{Giacomelli_2000} etc. 

Recently, a novel  promising method, the Out-of-time-Ordered Correlator (OTOC) has been put forward to study spatio-temporal chaos in extended systems \cite{Das_2018, Khemani_2018}. This quantity, denoted as $D(x,t)$, measures the growth (in time) and spread (in space) of a infinitesimal localized perturbation in the initial conditions of two copies of the system.  Usually the OTOC is presented in the form of a heat-map in space-time which has  light-cone like structures~\cite{Das_2018}. Such structures are described by a ballistic spread and growth of  perturbation, characterized by butterfly speed $v_b$ (essentially of the cone) and the Lyapunov exponent $\lambda$.

Although this has generated a lot of interest, the use of OTOC as a diagnostic in classical extended systems has been restricted to a very few cases, such as classical Heisenberg spin chain at infinite temperature \cite{Das_2018}, thermalised fluid obeying Galerkin-truncated inviscid Burgers equation \cite{Kumar_2019} and classical interacting spins on Kagome lattice \cite{Bilitewski_2018}. It is important to note that {\it most} of these works were on Hamiltonian systems. Studies in systems lacking a Hamiltonian structure, specially, in driven-dissipative systems is essentially unexplored. In this paper, we address spatio-temporal chaos in extended driven-dissipative system using Duffing chain as a platform. 

The idea of OTOC originates from the fascinating and well developed notion of Out-of-time-Ordered Commutator in quantum systems widely used to study scrambling of information and quantum chaos  \cite{Sekino_2008,Shenker_2014,Rozenbaum_2017,Kukuljan_2017,Bohrdt_2017,Lakshminarayan_2019}. This measures the generation (in space-time) of non-commutativity of otherwise initially commuting operators in extended quantum systems.  There have been recent works where Out-of-time-Ordered Commutators play a prominent role. For example, it has been used to understand the effect of dissipation in quantum systems \cite{Zhang_2019,Loga_2019}, to characterize thermal and Many Body Localized phases \cite{He_2017,Fan_2017}, to understand localization to delocalization transition in quasi-periodic systems (e.g.  
Aubry-Andr$\acute{\mathrm{e}}$ model) \cite{Ray_2018}, to study scrambling of information in both integrable and non-integrable models such as Sachdev-Ye-Kitaev
 model \cite{Polchinski_2016,Maldacena_2016}, 1-d quantum Ising spin chain \cite{Lin_2018}, 
Floquet-Frederickson-Anderson model \cite{Gopalakrishnan_2018}, disordered XY spin chain \cite{McGinley_2019} and to explore super-diffusive broadening of fronts in long-ranged power law interaction systems \cite{Chen_2019}.   
%
%
 
Despite this considerable work on extended quantum systems, as mentioned earlier, very little has been investigated in extended classical Hamiltonians and essentially nothing is explored in non-Hamiltonian systems. To address this lack of understanding, in this paper, we study spatio-temporal chaos in {\it driven dissipative} chain of coupled Duffing oscillators (DC) using OTOC. This is a rich nonlinear system which exhibits plethora of exciting complex dynamical phenomena.  In the context of investigating various intriguing phenomena like chaos, multivalued amplitude-response,  synchronization and chimera states, to name a few,  systems with single or few Duffing oscillators have been  extensively and successfully used as a major platform \cite{Duffing_1918,Ueda_1978,Ueda_1979,Ueda_1991,Stupnicka_1987,Englisch_1991,Kovacic_2011,Gottwald_1992,Wei_1997,Chabreyrie_2011, 
Kozlowski_1995,Kenfack_2003,Musielak_2005,Jothimurugan_2016,Wei_2011,Kapitaniak_1993,Lai_1994,Clerc_2018}. In addition Duffing oscillator can be used in various practical applications. For example, Duffing oscillator based encryption devices have been proposed for secure communication systems \cite{Zapateiro_2013,Murali_1993}. Duffing oscillators can be used in weak signal detection in various cases like fatigue damage in materials \cite{Hu_2003,Zhang_2017} and down-hole acoustic telemetry in oilfield exploration \cite{Liu_2011}.  Such broad applications of Duffing oscillators and progress in theory \cite{Kovacic_2011} as well as in experiments \cite{Murali_1993}, makes Duffing chain a natural test-bed for studying spatio-temporal chaos in 
extended driven dissipative classical systems - an area yet largely unexplored. Below we briefly summarize our main observations and findings (see also TABLE \ref{tab:table}).
 
\begin{figure}[H]
  \centering
  \subfigure[$\left\lbrace \bar{f}=0.30,\bar{\gamma}=0.15,\bar{\kappa}=1.0\right\rbrace$]{\includegraphics[scale=0.39]{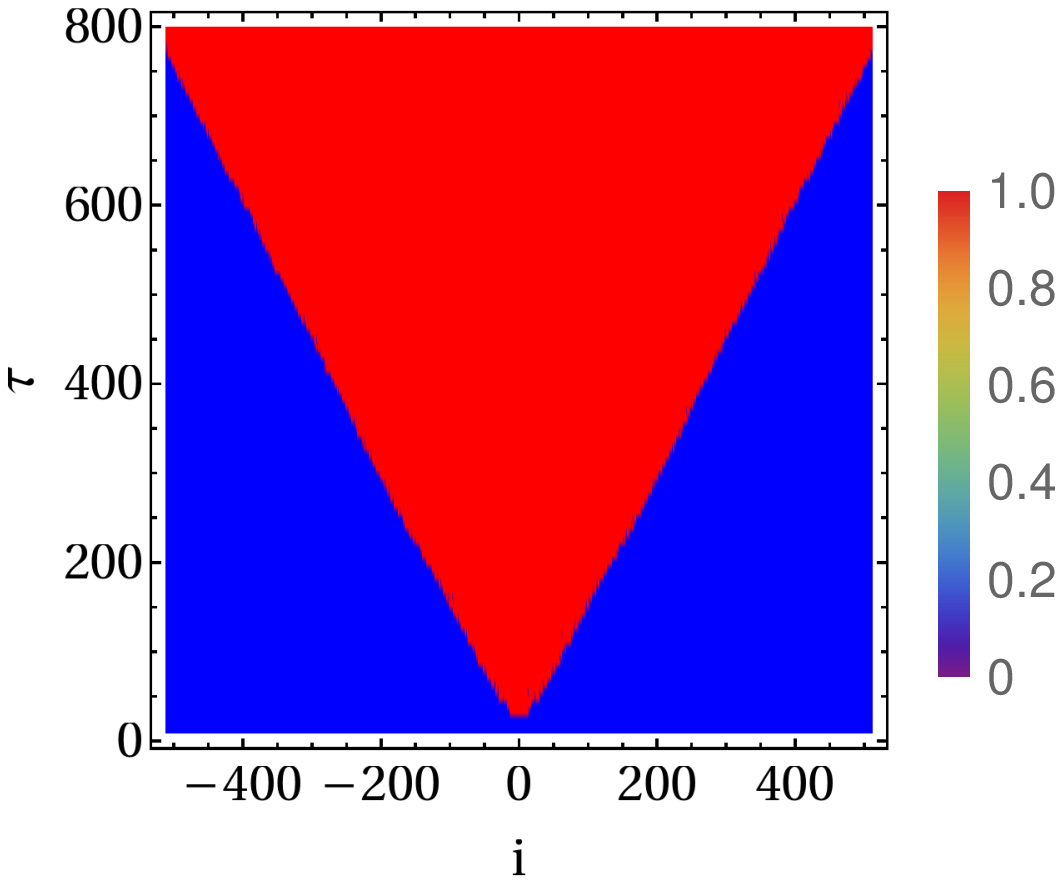}}\hfill
  \subfigure[$\left\lbrace \bar{f}=0.09,\bar{\gamma}=0.01,\bar{\kappa}=2.0 \right\rbrace$]{\includegraphics[scale=0.39]{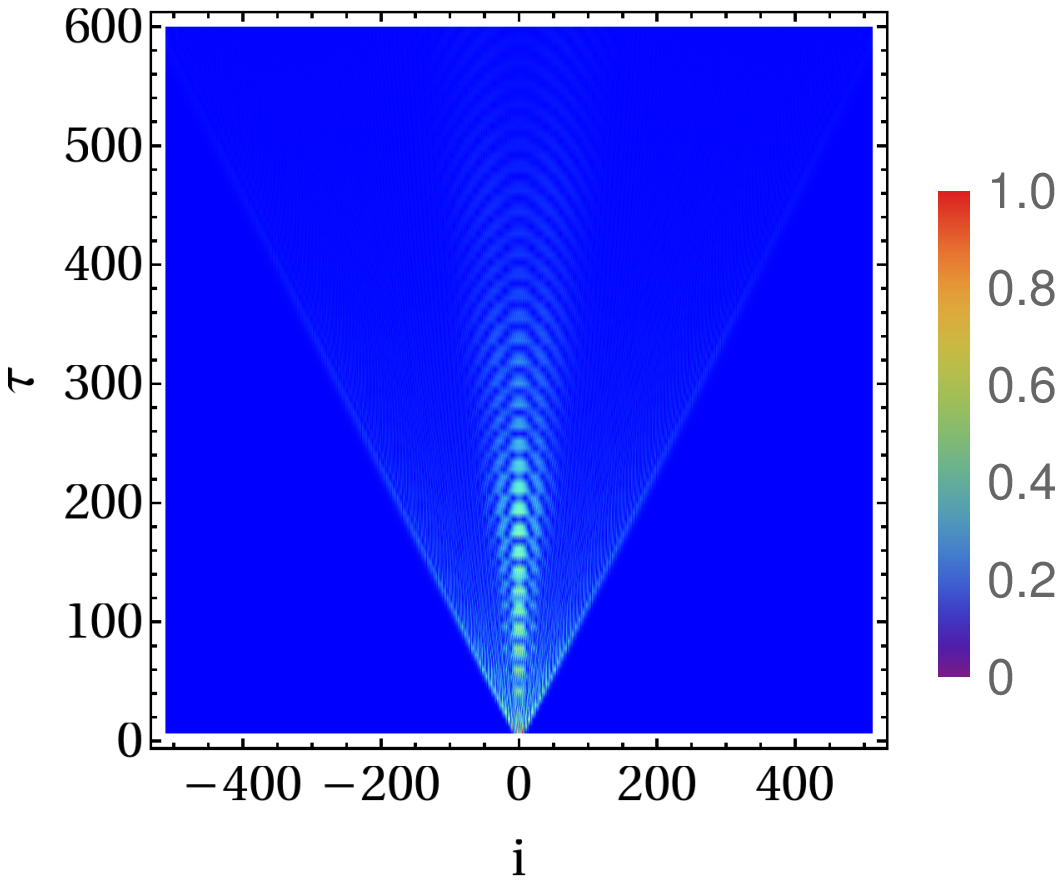}}\\
  \subfigure[$\left\lbrace \bar{f}=0.24,\bar{\gamma}=0.15,\bar{\kappa}=1.0 \right\rbrace$]{\includegraphics[scale=0.39]{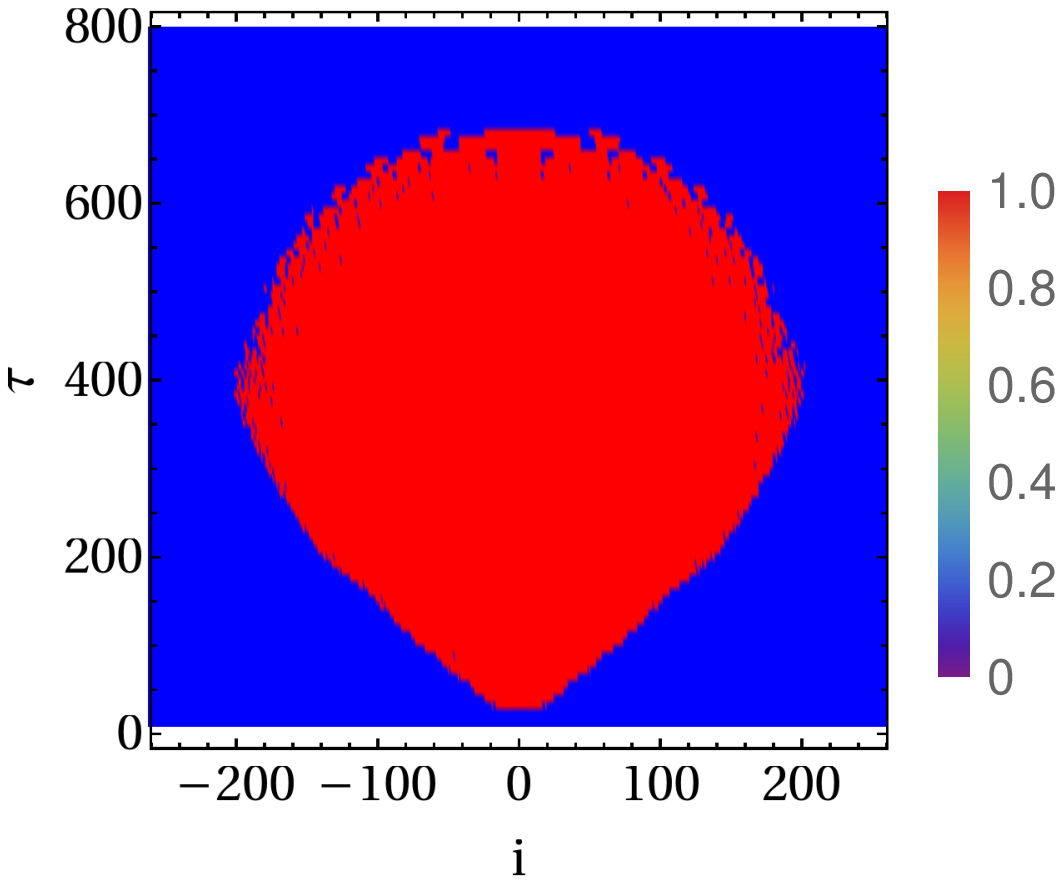}}\hfill
  \subfigure[$\left\lbrace \bar{f}=0.13,\bar{\gamma}=0.15,\bar{\kappa}=1.0 \right\rbrace$]{\includegraphics[scale=0.39]{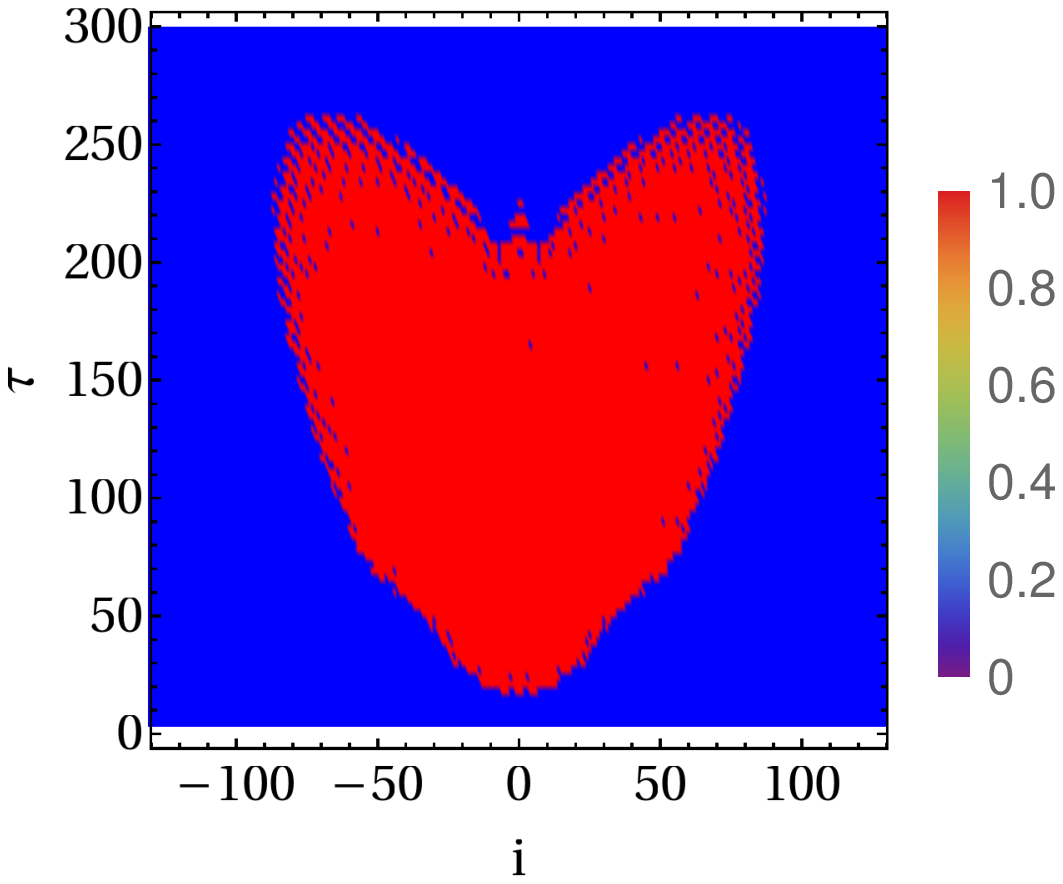}}
  \caption{Spatio-temporal spread of perturbation in a driven dissipative Duffing chain (DC) exhibits different dynamical regimes. The OTOC [see Eq.~(\ref{otoc})] for a DC of length $N=1025$ shows (a) ballistic spread and exponential growth forming {\it light cone} in {\it sustained chaos} regime, (b) short time ballistic spread and exponential decay creating an initial {\it light cone} that vanishes rapidly in {\it non-chaotic} regime (c) initial growth and ballistic spread followed by exponential decay and non-ballistic behavior, initial light-cone deforms into {\it balloon} shape in {\it transient chaos} regime, (d) similar qualitative behavior as (c), only the initial light-cone deforms into {\it butterfly} shape in {\it transient chaos} regime. Initially ($\tau=0$) the middle oscillator ($i=0$) is perturbed with $\epsilon=10^{-6}$.}
\label{heatmap}
\end{figure}

(i) We present OTOC as a remarkable diagnostics for demarcating various regimes of dynamical behaviors of a chain of coupled Duffing oscillators. The space-time heat-map plots of it show distinct patterns for the three dynamical regimes, called as {\it sustained chaos, transient chaos} and {\it non-chaotic} regimes (see Fig.~\ref{heatmap}). Although the existence of these three regimes was known from earlier works \cite{Umberger_1989}, a good diagnostic was missing. 

(ii) Given that the heat-map plots can be different from the conventional light-cone type maps (see Fig.~\ref{heatmap}), it necessitates generalizing the notion of concepts such as the butterfly velocities and the Lyapunov exponents. More precisely, we introduce the notion of {\it instantaneous butterfly speed} (IS) and use the generalized notion of finite time Lyapunov exponent (FTLE) \cite{Pazo_2016}. These notions proved to be key for understanding finite time behavior and transitions between different dynamical regimes.

(iii) We observe that in the sustained chaos regime, the growth of the perturbation measured in a frame moving with speed $v$ is exponential with a Lyapunov exponent $\lambda(v)$ dependent on $v$. Such velocity dependent Lyapunov exponent, known as VDLE \cite{Khemani_2018} or  convective LE \cite{Giacomelli_2000}, has been studied recently \cite{Das_2018,Khemani_2018} where it was observed that $\lambda(v)$ depends linearly on $v$ for $v \sim v_b$. In our case also, we observe such linear dependence. However, interestingly, the {\it detailed form} of the VDLE for DC has been observed to be different from what have been reported earlier for chaotic Hamiltonian systems \cite{Das_2018,Bilitewski_2018,Khemani_2018}.

(iv) In the transient chaos regime, the  OTOC grows initially (as a conventional light-cone), which is characterized by FTLE. After this initial dynamics, there is a simultaneous decrease in the FTLE at a specific time in all oscillators that have gained positive FTLE by this time. This effect is manifested in the corresponding OTOC heat map as emergence of complex geometrical shapes. We also find that once there is this decrease, the subsequent features can be quantitatively explained via analytical results from driven-dissipative harmonic chain.


(v) The variation of the IS and FTLE with tunable parameters exhibit several interesting features. With the continuous increase of the driving amplitude, the Duffing chain transits from non-chaotic to sustained chaos regime. This transition is interestingly preceded by appearance of intermittent transient chaos windows and sustained chaos points inside the non-chaotic regime. Deep inside the sustained chaos regime, the FTLE (that, in the large time limit, saturates to the conventional Lyapunov exponent) increases linearly with driving amplitude. In case of tuning the dissipation, stating from a chaotic regime, the FTLE decreases approximately linearly with increasing dissipation followed by an highly intermittent behavior with mixture of chaotic and periodic windows. In context of coupling, our investigation reveals that a chain of uncoupled Duffing oscillators in non-chaotic regime can be made to transit to chaotic regime only by tuning the coupling strength. Also, the IS exhibits a power law increase ($v_b \sim \kappa^{\sigma}$ with $\sigma=0.71$) with increasing coupling strength. Notably, the value of $\sigma$ for the DC is different from that of $\sigma=1/2$ in case of a driven dissipative harmonic chain (shown analytically in Appendix.~\ref{AI}). This indicates to the important role of nonlinearity on the speed of spatial spread of an initially localized perturbation. 

(vi) For the case of zero nonlinearity, i.e., for driven dissipative harmonic chain we present rigorous analytical results for the OTOC and VDLE (Appendix.~\ref{AI}). Results for OTOC are obtained in terms of Airy function and the effect of openness (dissipation) is elaborated. The behavior of VDLE is extracted.

\section{Model and tools (OTOC, IS, FTLE)}
\label{model}
\noindent
We consider a driven dissipative ring of $N$ Duffing oscillators, with nearest neighbor harmonic coupling where every oscillator is coherently driven by an external periodic force of frequency $\Omega$ and strength $f$. The equation of motion for the $i$-th oscillator with position $x_i(t)$ at time $t$ is given by
\be
\ddot{x}_i=k_0 x_i-\alpha x_i^3- \gamma \dot{x}_i+\kappa (x_{i+i}+x_{i-1}-2 x_i)+f \mathrm{cos}(\Omega t), \label{eqm1}
\ee
where $\alpha,\gamma,\kappa,k_0$ are nonlinearity, damping, harmonic coupling constant and spring constant respectively. We need $\alpha >0$ to ensure that the onsite potential is confining.  Also note that $\gamma >0$ to make sure that the system does not heat up. We restrict ourselves to $k_0>0$ to make the onsite potential double-well in nature.
For this  model, we aim to study the possibility of chaotic, transient and regular motions of this spatially extended chain of Duffing oscillators in the parameter space constituted by  $\left\lbrace f,\gamma, \kappa,\alpha,\Omega,k_0 \right\rbrace $  using OTOC as a tool.

To start with, it is important to note that using proper scaling it is possible to reduce the number of independent scaling  parameters. We define the following new variables 
\begin{align}
y_i=\sqrt{\alpha}~x_i,~\tau=\sqrt{k_0}~t, 
\end{align}
so that Eq. (\ref{eqm1}) gets transformed to
\be
\ddot{y}_i= y_i- y_i^3- \bar{\gamma} \dot{y}_i+\bar{\kappa} (y_{i+i}+y_{i-1}-2 y_i)+\bar{f}\mathrm{cos}(\bar{\Omega} \tau), \label{eqm}
\ee
where 
\begin{align}
\bar{\gamma}=\frac{\gamma}{\sqrt{k_0}},~
\bar{\kappa}=\frac{\kappa}{k_0},~
\bar{f}=\frac{f}{k_0} \sqrt{\alpha},~
\bar{\Omega}=\frac{\Omega}{\sqrt{k_0}}.
\label{s-parameters}
\end{align}
In order to explore different dynamical behaviors of the extended Duffing chain we study OTOC in this rich parameter space.

To measure OTOC, we start with two identical copies (I~\&~II)  of the same Duffing-chain with the only difference being an infinitesimal difference $\epsilon$ in the initial conditions at a chosen oscillator (say the middle one). We now let the two copies evolve independently according to Eq.~\eqref{eqm} and observe how the initial difference spread and grow in space-time which can be captured by OTOC [$D(i,\tau)$] defined as 
\be
D(i,\tau)=\frac{\big{|}y_i^I(\tau)-y_i^{II}(\tau)\big{|}}{\big{|} y_{\text{middle}}^I(0)-y_{\text{middle}}^{II}(0) \big{|}} = \frac{\big{|}y_i^I(\tau)-y_i^{II}(\tau)\big{|}}{|\epsilon|}, \label{otoc}
\ee
 This quantity measures the ratio of the deviation between the two copies for the $i$th oscillators at time $\tau$ to the deviation  $\epsilon$ for the middle oscillator at $\tau=0$.  Naturally, $D(i,\tau)$ 
captures information of both the temporal growth (or decay) and spatial spread of the initial deviation. To extract these information,  we define 
IS $v_b(\tau,D_{\text{th}})$ and FTLE $\lambda_i(\tau)$ from the OTOC  $D(i,t)$ as  
\begin{align}
 v_b(\tau,D_{\text{th}}) &=\frac{\sum_{i=1}^N\Theta[D(i,\tau)-D_{\text{th}}]}{\tau} \label{IS} \\
\lambda_i(\tau) &= \frac{\mathrm{ln}~D(i,\tau)}{\tau} . \label{ftle}
\end{align}
where $\Theta(x)$ is a step function. The IS $v_b(\tau,D_{\text{th}})$  in the above equation measures the number of oscillators (per unit time) that have gained deviations greater or equal to $\epsilon D_{\text{th}}$. On the other hand, FTLE  $\lambda_i(\tau)$ describes how the deviation at a particular oscillator grows or decays with time.

As mentioned earlier, the extended Duffing chain exhibits three different types of dynamical behavior, namely sustained chaos, transient chaos and non-chaotic behavior. It is exciting to see how IS and FTLE can characterize and distinguish between all these dynamical regimes.  In case   of sustained chaos, one would expect that, in the long $\tau$ limit,  $\lambda_i(\tau)$ will eventually saturate to some positive constant  $\lambda$ independent of $i$. Based on recent works on Hamiltonian systems~\cite{Das_2018,Bilitewski_2018}, the IS is also expected to approach a constant value (time independent), $v_b$ which is known as butterfly speed. On the other hand, in the non-chaotic (regular) regimes it is expected that $\lambda_i(\tau) \leq 0$ $\forall \tau,i$ and saturates to $\lambda \leq 0$ for large $\tau$. However, the notion of butterfly speed for the non-chaotic regime, strictly speaking, ceases to exist. None-the-less, one can, define a spreading speed in the $D_{\text{th}} \to 0$ limit. The transient regime exhibits intricate interplay between nonlinearity, dissipation and drive. This regime shows a crossover from chaotic dynamics to regular dynamics. This crossover is characterized by change in sign of FTLE from positive to negative. 
%

To explore these features, in the next section, we numerically compute IS and FTLE from OTOC and analyze  in detail how they can describe the three different dynamical regimes in DC.

\section{Numerical results}
\label{numerical}

In this section we numerically compute OTOC defined in Eq.~\eqref{otoc} in the $\epsilon\to 0$ limit. In this limit, one can in fact write an evolution equation for $\delta y_i(\tau)= y_i^{I} (\tau) -y_i^{II} (\tau)$ which to leading order in $\epsilon$ is given by 
\be
\frac{d^2 \delta y_i}{d\tau^2}=(1-3 y_i^2)\delta y_i- \bar{\gamma} \frac{d \delta y_i}{d\tau}+\bar{\kappa} (\delta y_{i+i}+\delta y_{i-1}-2 \delta y_i),
\label{eqm_lm}
\ee
where $y_i(\tau)$ present in the first term is obtained by solving Eq.~\eqref{eqm}. This term makes this equation a linear ODE with time dependent coefficient and this is the central cause of possible spread and growth of OTOC.  
To integrate Eqs.~\eqref{eqm}, \eqref{eqm_lm} numerically, we use fourth order Runge-Kutta(RK4) algorithm with time step $\Delta\tau=0.001$ and with initial conditions
\begin{align}
\begin{split}
y_i(0)&=y_0,~~~ \dot{y_i}(0)=0;\\
\delta y_i(0)&=\epsilon~ {\delta}_{i,\frac{N+1}{2}},~~~ \delta \dot{y}_i(0)=0, 
\end{split}
\label{init} 
\end{align}
for $i = 1,2... N$ where 
$y_0$ is a constant and $\delta_{i,j}$ is the usual Kronecker-Delta function. Note that the deviation $\delta y_i (0)$ is non-zero only at the middle site which can be thought of as an initial perturbation. 
For all numerical simulations,  we chose $\epsilon = 10^{-6}$ and  $\bar{\Omega}=1$. The space constituted by the other three parameters $\left\lbrace \bar{f},\bar{\gamma},\bar{\kappa} \right\rbrace$ are explored extensively to investigate the three different dynamical regimes. 

%
%
%

\begin{figure}[t]
 \centering \includegraphics[width=8 cm]{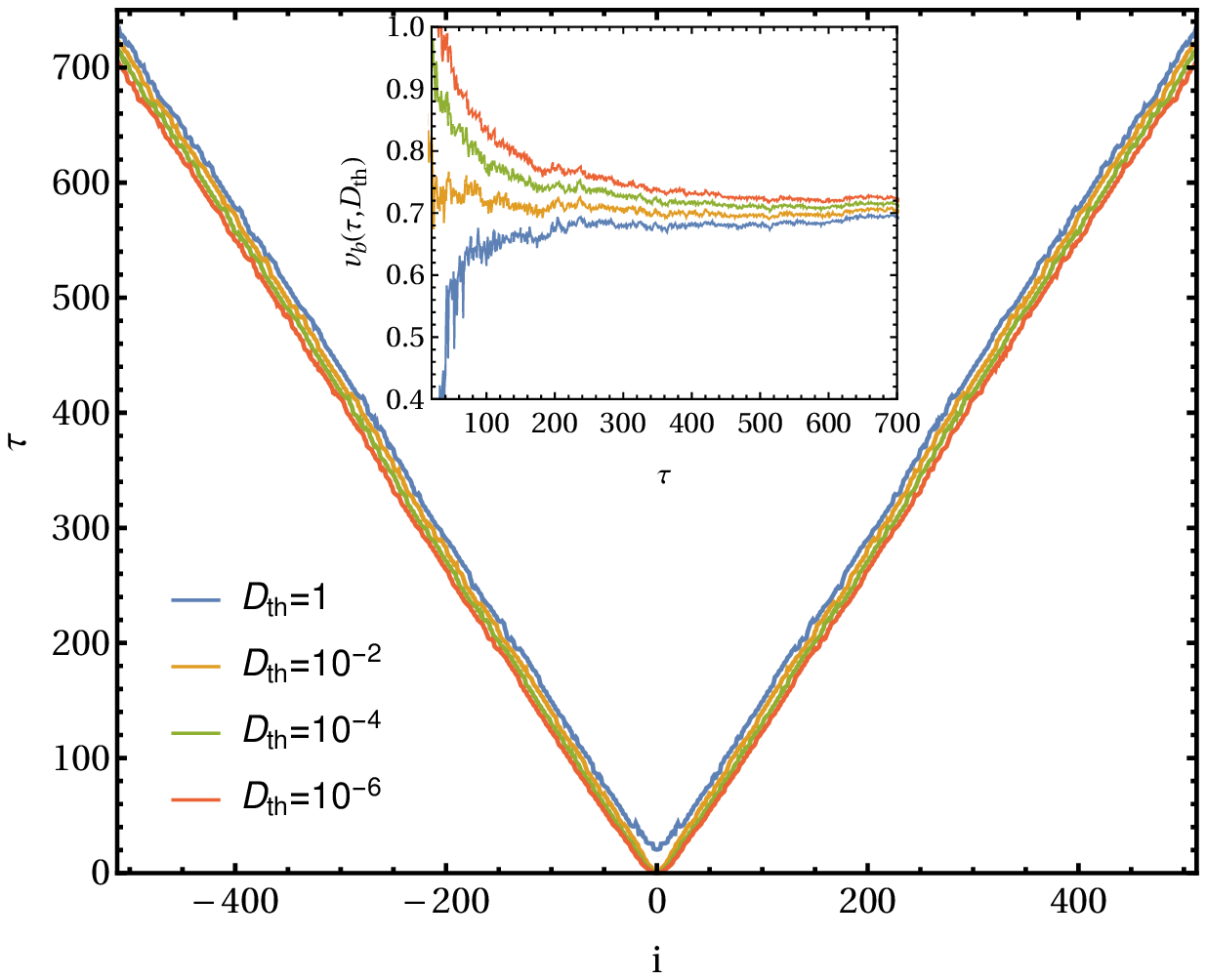} 
 \caption{\textit{sustained chaos:} $\left\lbrace \bar{f}=0.30,\bar{\gamma}=0.15,\bar{\kappa}=1.0\right\rbrace$. The main plot shows the boundaries of the OTOC light-cone for different values of threshold ($D_{\mathrm{th}}$) lie very close, implying the ballistic spread is independent of $D_{\mathrm{th}}$. In the inset plot,  the slopes [$v_b(\tau,D_\mathrm{th})$] of these boundaries computed from Eq. (\ref{IS}) are indeed observed to be independent of the value of $D_{\text{th}}$ and they saturate to a constant butterfly speed $v_b=0.7.$}
 \label{fig:sc_boundary}
\end{figure}

\subsection{Sustained chaos regime}
\label{sec:sc}
\begin{figure}[t]
  \centering \includegraphics[width=8 cm]{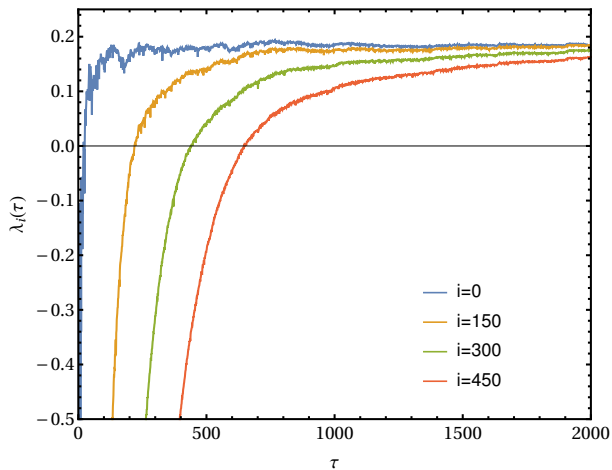}
  \caption{\textit{sustained chaos:} $\left\lbrace \bar{f}=0.30,\bar{\gamma}=0.15,\bar{\kappa}=1.0\right\rbrace$. FTLE [computed from Eq. (\ref{ftle})] for different oscillators ($i$) saturate to the same constant value, the conventional Lyapunov exponent $\lambda_i(\tau)=\lambda=0.18$ identifying the exponential growth of OTOC.}
\label{fig:ftle_sc}
\end{figure}

In this case, we  carefully choose the parameter values to be $\left\lbrace \bar{f}=0.30,\bar{\gamma}=0.15,\bar{\kappa}=1.0 \right\rbrace$ with  $y_0=0.3$ to observe sustained chaos regime in a DC of length $N=1025$.  In Fig.~\ref{heatmap}(a), we present the heat-map of $D(i,\tau)$ that exhibits a {\it light-cone} like structure implying ballistic propagation of perturbation along the chain. The speed of the propagation can, in principle, be obtained from the slope of the boundary between the dark and bright region of the heat map. Instead of using this method, we employ a more accurate method of determining the boundary line. 
At a given $\tau$ we find out the farthest oscillator $i$ from the middle in either direction such that $D(j,\tau)<D_{\text{th}}$ for $|j-(N+1)/2|>|i-(N+1)/2|$. 
We plot such boundaries for different $D_{\text{th}}$ in Fig.~\ref{fig:sc_boundary} and we observe that the slope of these boundary lines are independent of $D_{\text{th}}$. An equivalent way of extracting this speed is by computing the IS defined in Eq.~\eqref{IS} and this is plotted in the inset of Fig.~\ref{fig:sc_boundary} where we see that it saturates to $v_b = 0.7$. To measure the rate of growth of the perturbation, in Fig.~\ref{fig:ftle_sc}, we plot FTLE $\lambda_i(\tau)$ for different values of $i$. We observe that in the large $\tau$ limit FTLEs for all the oscillators reach the conventional Lyapunov exponent $\lambda$ which, for the parameter set $\left\lbrace \bar{f}=0.30,\bar{\gamma}=0.15,\bar{\kappa}=1.0 \right\rbrace$, has the value $\lambda=0.18$. This implies that the initial  perturbation localized at the middle point grows exponentially with time and spreads to all the oscillators making the whole DC chaotic. The fact that the FTLEs for all the oscillators reach $\lambda>0$ and stay there, ensures that the DC sustains its chaotic behavior indefinitely. 

\begin{figure}[t]
 \centering \includegraphics[width=8 cm,height= 6cm]{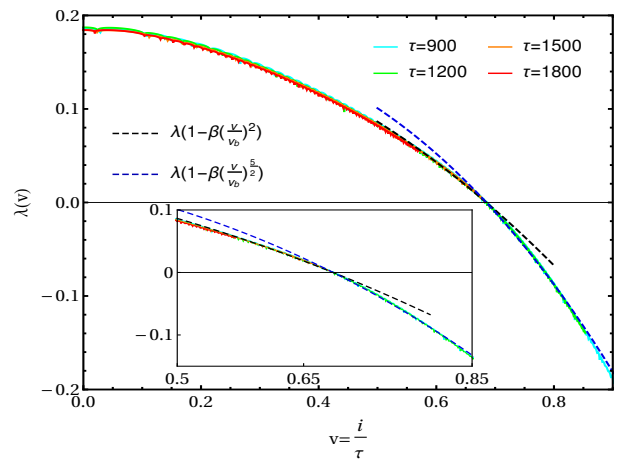}
\caption{\textit{sustained chaos:} $\left\lbrace \bar{f}=0.30,\bar{\gamma}=0.15,\bar{\kappa}=1.0\right\rbrace$. VDLE  [see Eq. (\ref{vdle})] exhibits a linear dependence $\lambda(v)\sim (v-v_b)$ near $v\approx v_b$. However, for $v\lesssim v_b$ and $v\gtrsim v_b,$ $\lambda(v)$ falls off with different exponents, $\lambda(v)=\lambda \big[1-\left(\frac{v}{v_b}\right)^\nu \big]$ with $\nu=2$ and  $\nu=\frac{5}{2}$ respectively. Here $v_b=0.69$ and $\lambda=0.18$.}
\label{muv-sc-1}
\end{figure}

The facts that the OTOC  grows exponentially and spreads ballistically suggest that OTOC has the following scaling form
\begin{align}
\lim_{\tau \to \infty}\frac{\ln D(i,\tau)}{\tau }= \lim_{\tau \to \infty} \lambda_i(\tau)= \lambda \left( {i}/{\tau}\right)=\lambda(v).
\label{vdle}
\end{align}
which we verify numerically in Fig.~\ref{muv-sc-1} via excellent data collapse. 
Existence of such scaling function implies that the perturbation observed in a frame moving with a velocity $v=i/\tau$ also grows/decays exponentially with a velocity dependent Lyapunov exponent (VDLE) $\lambda(v)$.  Concepts similar to VDLE, has been introduced earlier in the context of finite group velocity (Lieb-Robinson bound) in quantum spin systems with finite range interactions \cite{Lieb_1972}. These velocity dependent exponents, also known as convective Lyapunov spectrum \cite{Giacomelli_2000},  have been reported in studies of coupled map lattices \cite{Deissler_1984,Kaneko_1986}, complex Ginzburg-Landau equation \cite{Deissler_1987}, FPU chain \cite{Giacomelli_2000}, classical Heisenberg spin chain \cite{Das_2018}, interacting spins on Kagome lattice \cite{Bilitewski_2018} etc. 

\begin{figure}[t]
 \centering \includegraphics[width=8 cm,height= 6cm]{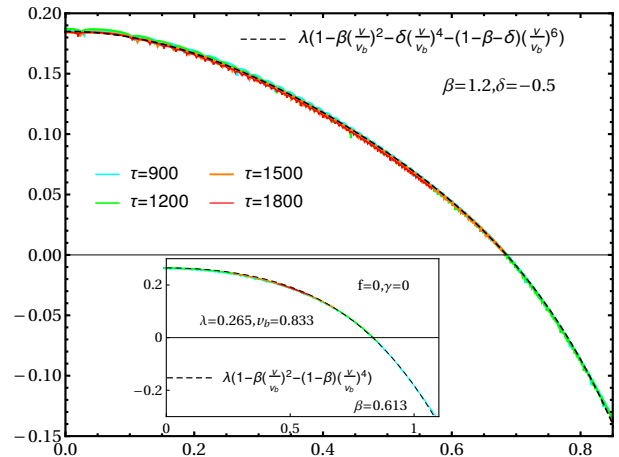}
\caption{\textit{sustained chaos:} $\left\lbrace \bar{f}=0.30,\bar{\gamma}=0.15,\bar{\kappa}=1.0\right\rbrace$. For any value of $v$ (not necessarily in the neighborhood of $v_b$), the VDLE for DC follows the functional form $\lambda(v)=\left[1-\beta(v/v_b)^2-\delta(v/v_b)^4-(1-\beta-\delta)(v/v_b)^6\right]$ with $\beta=1.2$ and $\delta=-0.5$. In the inset, the corresponding Hamiltonian counterpart with $\bar{f}=0$ and $\bar{\gamma}=0$ is plotted and observed that the VDLE in this case has a different expression given by $\lambda(v)=\left[1-\beta(v/v_b)^2-(1-\beta)(v/v_b)^4\right]$ with $\beta=0.613$.}
\label{muv-sc-2}
\end{figure}

Interestingly, a universal framework for describing exponential growth or decay of OTOCs in classical, semi-classical and large-$N$ systems in terms of VDLE has recently been discussed in Ref.~\cite{Khemani_2018} and possible functional forms of $\lambda(v)$ for $v\sim v_b$ have been proposed. In particular, for chaotic classical systems, it has been analyzed that $\lambda(v)$ continuously approaches to zero both from inside and outside the light-cones as $\lambda(v)\sim |v-v_b|$. Such linear behavior has been verified for Hamiltonian systems \cite{Das_2018}, {\it e.g.} 
in a classical Heisenberg chain \cite{Das_2018} where it has been observed that  $\lambda(v) =\lambda\left[1-(v/v_b)^2\right]$ . All these results and discussions are mostly restricted to Hamiltonian systems. Therefore one ponders as to how VDLE $\lambda(v)$ would behave for driven dissipative systems.

Motivated by the observations $\lambda(v) =\lambda\left[1-(v/v_b)^2\right]$ for $v\approx v_b$ in classical spin chains \cite{Das_2018,Bilitewski_2018}, one could ask if the same relation also holds for other models. In particular, the situation for  driven dissipative system is even more elusive. None-the-less, we use this form of $\lambda(v)$ for DC and obtain the following exponents (see Fig.~\ref{muv-sc-1}), 
\begin{align}
\label{slopessc}
\lambda(v)=
\begin{cases}
\lambda\left[1-(v/v_b)^2\right],~\text{for}~v\lesssim v_b \\
\lambda\left[1-(v/v_b)^{5/2}\right],~\text{for}~v \gtrsim v_b. \\
\end{cases}
\end{align}
So, as seen in the context of Hamiltonian systems \cite{Khemani_2018,Das_2018}, the function $\lambda(v)$ in our case goes to zero linearly as $v$ approaches $v_b$. But interestingly,  we find that the slopes (Eq.~\ref{slopessc}) of this linear behavior  are dramatically different for $v<v_b$ and $v>v_b$. It is natural to ask if the VDLE for DC has some single functional form that holds for both $v<v_b$ and $v>v_b$. In this regard we find, 
\bea
\lambda(v)&=&\left[1-\beta(v/v_b)^2-\delta(v/v_b)^4-(1-\beta-\delta)(v/v_b)^6\right],\cr
\beta&=&1.2, ~~\delta=-0.5,
\label{eq:vdle_sc}
\eea
which is clearly depicted in Fig.~\ref{muv-sc-2}. Note that the coefficient in the last term of Eq.~\ref{eq:vdle_sc} is fixed because of the constraint $\lambda(v=v_b) = 0$.
One might ask, if the different exponent values observed in case of the DC in comparison to the Hamiltonian system in Ref.~\cite{Das_2018}, is arising due to the presence of drive and dissipation. To answer this, in the inset of Fig.~\ref{muv-sc-2}, we present the behavior of $\lambda(v)$ for $\bar{f}=0$ and $\bar{\gamma}=0$. We observe that still it deviates from the behavior of VDLE in case of Heisenberg spin chain in Ref.~\cite{Das_2018} and has the functional form $\lambda(v)=\left[1-\beta(v/v_b)^2-(1-\beta)(v/v_b)^4\right]$ with $\beta=0.613$. 

 The facts that the VDLE (i) can behave differently for different Hamiltonian systems and (ii) the introduction of drive and dissipation has further significant impact on the behavior of $\lambda(v)$ are interesting observations and require further exploration.


\subsection{Non-chaotic regime}
\label{sec:nc}
\begin{figure}[t]
 \centering \includegraphics[width=8 cm]{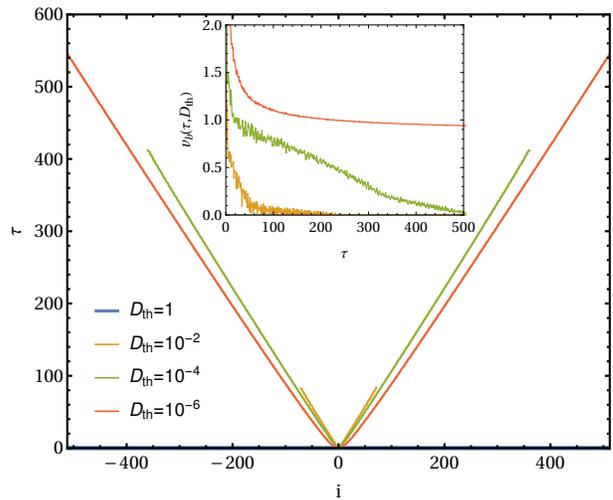} 
 \caption{\textit{non-chaotic:} $\left\lbrace \bar{f}=0.09,\bar{\gamma}=0.01,\bar{\kappa}=2.0 \right\rbrace$. The main plot shows the boundaries of the OTOC here strongly depend on $D_{\mathrm{th}}$, smaller the $D_{\mathrm{th}}$ value, larger is the light-cone boundary. The slopes (IS) of the corresponding boundaries, computed from Eq. (\ref{IS}), have well defined values only in $D_{\mathrm{th}}\to 0$ limit. In particular $v_b(\tau)=0.93$ for $D_{\mathrm{th}}=10^{-6}$ whereas $v_b(\tau)\approx0$ for $D_{\mathrm{th}}=1$. }
\label{fig:nc_vb}
\end{figure}

 The DC possess a non-chaotic regime characterized by non-growing OTOC. In Fig.~\ref{heatmap}(b) we give heat-map of OTOC for $\left\lbrace \bar{f}=0.09,\bar{\gamma}=0.01,\bar{\kappa}=2 \right\rbrace$ in a DC of length $N=1025$. In this map we find there is a light-cone like structure but importantly the boundary separating the regions inside and outside the cone ceases to exist at larger times. Therefore in this regime, strictly speaking the propagation speed defined in Eq.~\eqref{IS} is defined only in the $D_{\mathrm{ th}}\to 0$ limit.  This is seen in Fig.~\ref{fig:nc_vb} where we plot the boundary measured with different values of  $D_{\mathrm{th}}$ and we find that smaller the threshold larger the length of the boundary (implying that further oscillators feel smaller amount of perturbation). Hence the slope of the boundary gets a well defined value for propagation speed as $D_{\mathrm{th}} \to 0$. Same value is also obtained from direct computation of the propagation speed from Eq.~\eqref{IS} for a very small $D_{\mathrm{th}}$ as presented in the inset plot of Fig.~(\ref{fig:nc_vb}). Note that the velocity obtained from the slope and from Eq.~\eqref{IS} may be different for finite $D_{\mathrm{th}}$ but they match in the limit $D_{\mathrm{th}} \to 0$.

\begin{figure}[t]
  \centering \includegraphics[width=8 cm]{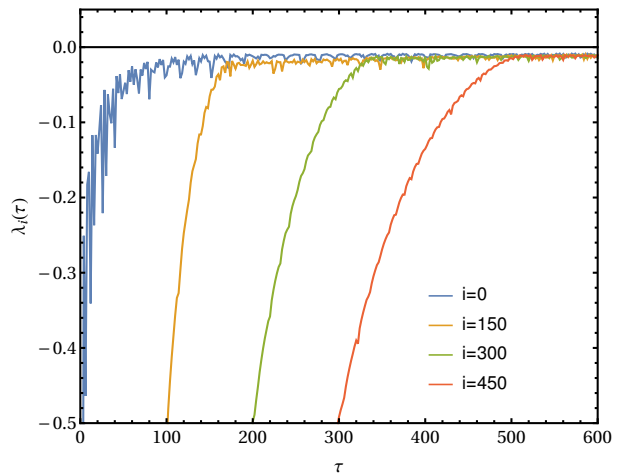}
  \caption{\textit{non-chaotic:} $\left\lbrace \bar{f}=0.09,\bar{\gamma}=0.01,\bar{\kappa}=2.0 \right\rbrace$. FTLE [computed from Eq. (\ref{ftle})] for different oscillators ($i$) saturate to the same negative constant, the conventional Lyapunov exponent $\lambda_i(\tau)=\lambda=-0.01$ identifying the exponential decay of OTOC.}
\label{fig:lvst_nc}
\end{figure}

In Fig.~\ref{fig:lvst_nc} we plot $\lambda_i(\tau)~vs.~\tau$ for different oscillators and we observe that they all saturate to a negative value, because dissipation dominates in this regime. Mathematically, $\lambda_i(\tau\rightarrow \infty)=\lambda_0 <0 ~~\forall i$. 

\begin{figure}[t]
 \centering \includegraphics[width=8 cm]{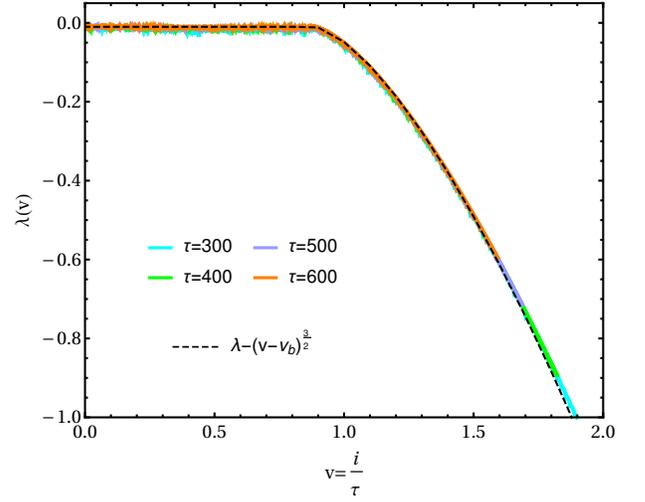}
\caption{\textit{non-chaotic:} $\left\lbrace \bar{f}=0.09,\bar{\gamma}=0.01,\bar{\kappa}=2.0 \right\rbrace$.  $\lambda(v)$ is always negative in the non-chaotic case. It is almost constant ($\lambda(v)=-0.01$) inside the cone ($v\leqslant v_b=0.886$) and falls as $(\lambda(v)-\lambda)\sim -(v-v_b)^{3/2}$ outside the cone ($v\geqslant v_b$).}
\label{fig:muv-nc-1}
\end{figure}

Motivated by our findings regarding VDLE in the sustained chaos case, we, in this case, explore how does $\lambda(v)$ scale with respect to $\tau$ and behave as function of $v=i/\tau$.
To investigate this, we present the corresponding numerical results in Fig.~\ref{fig:muv-nc-1} where we plot $\lambda(v)$ [as defined in Eq.~\eqref{vdle}] as a function of $v$. There, along with excellent data collapse at different time, we observe that 
\begin{align}
\lambda(v)=
\begin{cases}
\lambda,~\text{for}~v<v_b \\
\lambda-(v-v_b)^{\frac{3}{2}},~\text{for}~v>v_b, \\
\end{cases}
\label{eq:muv_nc}
\end{align}
where $\lambda_0=\lambda$. 

At this point it is worth noting that same behavior for the VDLE has been recently reported in  Ref.~\cite{Khemani_2018} for non-chaotic non-interacting Hamiltonian systems. 
In the non-chaotic regime, the dynamics in our problem becomes essentially a linear harmonic chain (non-interacting) because the particles 
execute small oscillations around the minima of the double well potential {\it i.e.} $y_i^2 \approx 1$. As a result the equation for 
the perturbation $\delta y_i$ in Eq.~\eqref{eqm_lm} becomes 
 \be
 \frac{d^2 \delta y_i}{d\tau^2} =-k_0 \delta y_i - \bar{\gamma} \frac{d \delta y_i}{d\tau} + \bar{\kappa} (\delta y_{i-1}+\delta y_{i+1}-2 \delta y_i),
 \label{hc_1}
 \ee
with $k_0=2$ where we have neglected the cubic term due to its smallness. In what follows, we demonstrate analytically that above equations of motion of a chain of coupled harmonic oscillators (HC) exhibits the behavior in Eq.~\eqref{eq:muv_nc} even in presence of dissipation for arbitrary $k_0>0$. It is also to be noted that although the original particle dynamics in Eq.~\eqref{eqm} is subjected to both drive and dissipation, the dynamics of the perturbation becomes insensitive to the drive because $y_i^2 \approx 1$ at all times.

Writing the general solutions for $\delta y_i$ in Eq.~\eqref{hc_1} exactly and using them in Eq.~\eqref{otoc} we obtain the following expression for OTOC in HC (see Appedix~\ref{AI} for details),
\bea
 && D(i,\tau)=\frac{\mathrm{e}^{-\bar{\gamma} \tau/2}}{N}  \sum_{j=1}^{N} \left[\mathrm{cos}\left(\frac{2 \pi i j}{N}-\Delta_j \tau\right) \right. \cr 
 &&~~~+ \left. \frac{\bar{\gamma}}{2\Delta_j}\mathrm{sin}\left(\frac{2 \pi i j}{N}-\Delta_j \tau\right)\right],
\label{hc_2}
\eea
where $ \Delta_j= \sqrt{4\bar{\kappa}~\mathrm{sin}^{2}\left(\frac{\pi j}{N}\right)+ k_0 -\left(\frac{\bar{\gamma}}{2}\right)^2}$. For a spatially extended large system, in the limit $N\rightarrow\infty,$ one can take the continuum limit of Eq. (\ref{hc_2}) by letting $\frac{\pi j}{N}=q$ so that
\bea
 && D(i=v\tau,\tau) = \frac{e^{-\frac{\bar{\gamma} \tau}{2}}}{\pi} \times \cr && \int_{0}^{\pi} \mathrm{d}q \left[\mathrm{cos}\left(2\tau(qv-\frac{1}{2}\Delta_q)\right)  -\frac{\bar{\gamma}}{2\Delta_q}\mathrm{sin}\left(2\tau(qv-\frac{1}{2}\Delta_q)\right)\right] \nonumber
\eea
 where $\Delta_q=\sqrt{2 \bar{\kappa}}\sqrt{1+\eta-\mathrm{cos}(2q)}$ with $\eta=\frac{k_0-(\bar{\gamma}/2)^2}{2\bar{\kappa}}$. A saddle point approximation of the integrand yields 
 (see again Appendix~\ref{AI} for details),
 \begin{eqnarray}
 D(v\tau,\tau)&=& 
 \begin{cases}
& \frac{2 e^{-\frac{\bar{\gamma} \tau}{2}} g(q^\ast)}{(4v_b \tau)^{1/3}} \mathrm{Ai}(z),~~~~\text{for}~v \gtrsim v_b \\
&\frac{2 e^{-\frac{\bar{\gamma} \tau}{2}} g(q^\ast)}{(4v_b \tau)^{1/3}} \mathrm{Ai}(-z),~~\text{for}~v \lesssim v_b 
\end{cases}
\label{D_vt_1} \\
\text{with} && z=\frac{2^{\frac{1}{3}}|v_b-v|\tau^{\frac{2}{3}}}{v_b^{\frac{1}{3}}} >0. \nonumber 
\label{hc_5}
\end{eqnarray}
where $\mathrm{Ai}(z)$ is the Airy function. Here, $v_b= \sqrt{\bar{\kappa}}~\sqrt{1+\eta-\sqrt{(1+\eta)^2-1}}$ and  $g(q^\ast,\tau)=\mathrm{cos}[2\tau(vq^\ast-\frac{1}{2}\Delta_{q^\ast})]-\frac{\bar{\gamma}}{2\Delta_{q^\ast}} \mathrm{sin}[2\tau(vq^\ast-\frac{1}{2}\Delta_{q^\ast})]$ with $q^\ast$ given by the solution of  $\mathrm{cos}(2q^\ast) = (1+\eta) - \sqrt{(1+\eta)^2-1}$.

In the limit $\tau\rightarrow\infty$, using the large $z$ asymptotic of Airy functions,  we have
\begin{align}
D(v\tau,\tau)=
\begin{cases}
\frac{ \hat{g}(q^\ast,\tau)}{2\sqrt{\tau}}~~ e^{-\frac{\bar{\gamma}}{2}\tau-\frac{2^{\frac{5}{2}}}{3\sqrt{v_b}}\tau(v-v_b)^{\frac{3}{2}}},~~~~~~~~~~~~v>v_b \\
\frac{e^{-\frac{\bar{\gamma} \tau}{2}} \hat{g}(q^\ast,\tau)}{\sqrt{\tau}}~ \mathrm{sin}\left[\frac{\pi}{4}+\frac{2^{\frac{5}{2}}\tau}{3\sqrt{v_b}}(v-v_b)^{\frac{3}{2}}\right],v<v_b 
\end{cases}
\label{hc_18}
\end{align}
where $\hat{g}(q^\ast,\tau)=\frac{g(q^\ast,\tau)}{2^{-\frac{3}{4}}\sqrt{\pi}(v-v_b)^{\frac{1}{4}}v_b^{\frac{1}{4}}}$. Notably in Eq. (\ref{hc_18}), apart from the explicit exponential dependence of the OTOC on dissipation (as $e^{-\bar{\gamma} \tau/2}$), $D(v\tau,\tau)$ depends on $\gamma$ through $v_b(\bar{\gamma})$ and $q^\ast(\bar{\gamma})$ in a nontrivial way.  We find that more the dissipation ($\gamma$), more is the butterfly velocity. This might seem counter-intuitive at first. Note that this measures how far a perturbation (however small it may be) can reach rather than the magnitude of perturbation. In fact the magnitude of the perturbation reached is suppressed exponentially with time. 
From Eq. (\ref{hc_18}), it is easy to see that VDLE $\lambda(v)$ defined in Eq. (\ref{vdle}) is given by Eq.~\eqref{eq:muv_nc}.

It is quite intriguing that, although the DC being a non-Hamiltonian nonlinear system the VDLE for DC, in the non-chaotic regime, exhibits  same exponents as reported for non-interacting integrable Hamiltonian systems \cite{Khemani_2018}.

\subsection{Transient chaos regime}
\label{sec:tc}
In the previous sections \ref{sec:sc} and \ref{sec:nc} we have observed that DC can exhibit sustained chaos or non-chaotic behavior depending on the choices of parameter values $\left\lbrace \bar{f},\bar{\gamma},\bar{\kappa}\right\rbrace$. The sustained chaos scenario is described by OTOC growing exponentially and spreading ballistically.
On the other hand, the non-chaotic regime is characterized by OTOC always decaying exponentially and  spreading ballistically at short time. In the sustained chaos regime, the FTLE starting from a negative value grows and finally saturates to a positive constant value  whereas in the non-chaotic regime the FTLE always remains negative.

In this section we demonstrate that by choosing the parameters carefully, one can observe a dynamical crossover from an exponentially growing and spreading OTOC ( similar to sustained chaos) regime  to a non-growing and  non-spreading OTOC (similar to non-chaotic) regime as time progresses. This interesting temporal crossover stems from the crucial presence of both drive and dissipation and, is manifested by un-conventional heat maps of OTOC as shown in Figs.~(\ref{heatmap}c) and (\ref{heatmap}d).
The existence of such a transient regime is far from obvious and has not been reported in generic Hamiltonian systems. In the context of DC (non-Hamiltonian), hints on existence of such regimes has been reported in Ref.~\cite{Umberger_1989} based on the observations of trajectories of the oscillators.  Using diagnostics based on OTOC and FTLE, our study reveals that this transient regime can be well characterized and contains in it a zoo of features as described below.
\begin{figure}[t]
 \centering \includegraphics[width=8 cm]{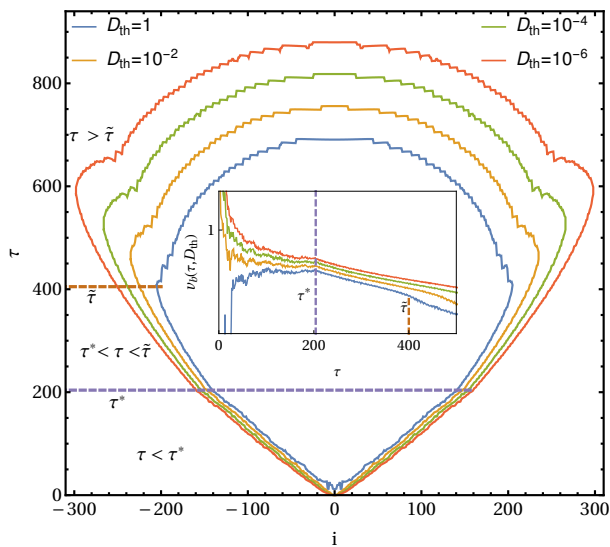}
\caption{\textit{transient chaos:} $\left\lbrace \bar{f}=0.24,\bar{\gamma}=0.15,\bar{\kappa}=1.0 \right\rbrace$. The main plot shows the boundaries of the OTOC for different values of $D_{\mathrm{th}}$. Initially up to $\tau\approx204,$ the ballistic spread forms light-cones and the boundaries are almost independent of $D_{\mathrm{th}}$. This is followed by a sudden change (marked by the dotted line) in the slopes of the boundaries marking the transit from light-cone to balloon shaped OTOC. The boundaries of these late-time balloon shapes are significantly different for different $D_{\mathrm{th}}$.  The inset plots of the IS (slope of the boundaries), computed from Eq. (\ref{IS}), show the existence of a constant speed (marked by the dotted line) up to $\tau\approx204$ after which a change in slope occurs indicating non-ballistic behavior of OTOC.}
\label{fig:tc_boundary}
\end{figure}


By optimum choice of parameters one can ensure to be in the transient chaos regime. As a sample example, we choose
 $\left\lbrace \bar{f}=0.24,\bar{\gamma}=0.15,\bar{\kappa}=1.0 \right\rbrace$ with  $y_0=0.3$  in the DC of length $N=1025$. The heat-map corresponding to these parameters in Fig.~\ref{heatmap}(c) shows that there is an initial time window ($0 \leq \tau < \tau^*$) in which the DC shares similarities with that of a chaotic system, characterized by light-cone like structure with sharp boundaries with a certain slope. 
There is a sudden behavioural change at $\tau = \tau^*$ after which the slope starts being time dependent thereby creating a sharp corner at $\tau = \tau^*$. This heat-map continues to spread however  with a time dependent speed till some time $\tilde{\tau}$ after which it stops spreading further. This rich beavior naturally demands a carefully analysis of the boundary of the heat map. In Fig.~\ref{fig:tc_boundary} we plot this boundary for different $D_{\mathrm{th}}$ values. Within the light-cone like structure ($\tau<\tau^*$), boundaries seem to converge for $D_{\mathrm{th}} \to 0$. However, for $\tau > \tau^*$, the boundaries depend on $D_{\mathrm{th}}$, although the qualitative features of them remain same (see Fig.~\ref{fig:tc_boundary}).  It is interesting to note that while $\tilde{\tau}$ is dependent on $D_{\rm{th}}$, $\tau^*$ is not. 
The existence and meaning of $\tilde{\tau}(D_{\rm{th}})$ can be understood best from the study of FTLE which we provide in the next paragraph. 
It is worth mentioning that these boundary features can be equivalently demonstrated by plotting IS versus $\tau$ for different $D_{\mathrm{th}}$ obtained from Eq.~\eqref{IS} as shown in the inset of Fig~\ref{fig:tc_boundary}. Note that the IS starts decreasing with time after  $\tau =\tau^{*}$. 
%

\begin{figure}[t]
  \centering \includegraphics[width=8 cm]{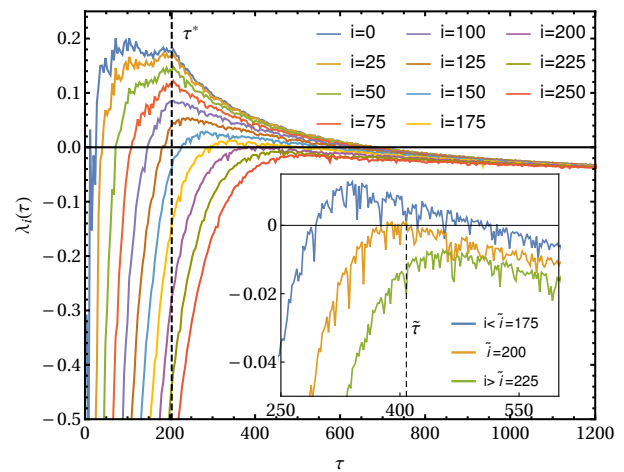}
 \caption{\textit{transient chaos:} $\left\lbrace \bar{f}=0.24,\bar{\gamma}=0.15,\bar{\kappa}=1.0 \right\rbrace$. 
 Initially FTLE $\lambda_i(\tau)<0 ~\forall i$. With time, FTLE for some oscillators (e.g. $i=0,50,100$) become positive (indicating chaos) whereas $\lambda_i(\tau)$ for $i=225,250$ etc. always remain negative (strictly non-chaotic). But the increase in $\lambda_i(\tau)$ for a bunch of chaotic oscillators (from $i=0$ to $i=100$ in this figure) stops near $\tau\approx204$ and suffers a  {\it sudden simultaneous} decrease (marked by the dotted line) - this is the same point at which change of slope in the light-cone and IS is observed in Fig.~\ref{fig:tc_boundary}. Finally, $\lambda_i(\tau)$ for all the oscillators (both transiently chaotic and non-chaotic) saturate to the negative constant $\lambda=-0.06$ indicating long time non-chaotic behavior.}
\label{fig:tc_ftle}
\end{figure}

To investigate the reason behind this sudden change in the slope as well as IS, we plot FTLE for different $i$ as a function of $\tau$ in Fig.~\ref{fig:tc_ftle} for $D_{\rm{th}}=1$. We observe that for $\tau < \tau^*$, FTLE for all the oscillators starts increasing with time.  Oscillators which are within the light-cone achieve positive values for the FTLE by this time. Remarkably, at $\tau =\tau^*$, FTLE of all these oscillators simultaneously starts decreasing. This is manifested by the sharp corner at $\tau=\tau^*$ of the heat map [see Fig.~\ref{heatmap}(c)]. Consequently, after this time the rate of spreading of the heatmap starts decreasing and at $\tau=\tilde{\tau}(D_{\rm{th}})$ it stops spreading as mentioned 
earlier. For a chosen $D_{\rm{th}}$ there exists an oscillator $\tilde{i}(D_{\rm{th}})$ whose FTLE barely touches zero from below at time $\tilde{\tau}(D_{\rm{th}})$ and remains negative after $\tilde{\tau}(D_{\rm{th}})$ as shown in the inset of Fig.~\ref{fig:tc_ftle} for $D_{\rm{th}}=1$  where $|\tilde{i}(1)|=200$. The oscillators with index $|i|>|\tilde{i}(D_{\rm{th}})|$ never achieve a positive FTLE suggesting that these oscillators never gain the initial perturbation  given at the middle ($0$-th) oscillator. 

\begin{figure}[t]
  \centering
  \subfigure[]{\includegraphics[scale=0.39]{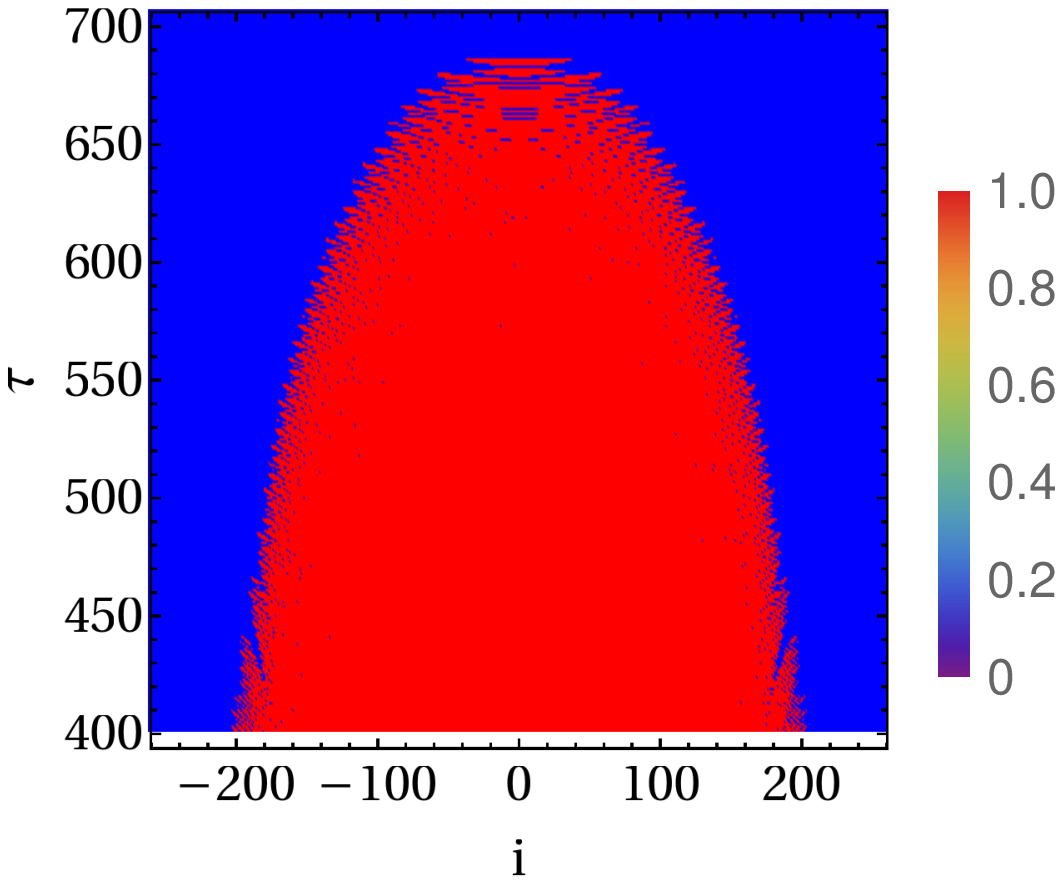}}\hfill
  \subfigure[]{\includegraphics[scale=0.39]{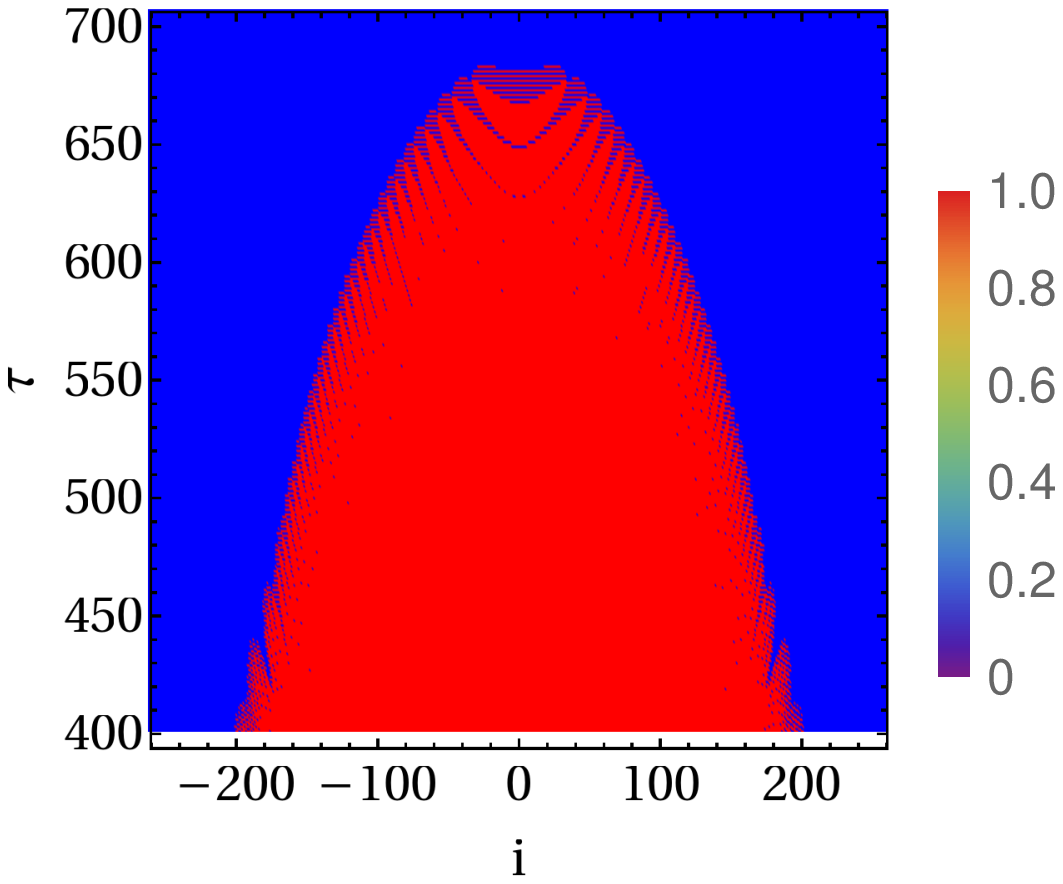}}
  \caption{\textit{transient chaos:} $\left\lbrace \bar{f}=0.24,\bar{\gamma}=0.15,\bar{\kappa}=1.0 \right\rbrace$.  (a) Transient chaos using original on-site double well potential (b) Transient chaos by approximating the double well potential by harmonic potential}
\label{fig:tc_harmonic}
\end{figure}

Once we cross the time scale $\tilde{\tau}$ the system starts behaving like non-chaotic regime which can be effectively described by a driven dissipative Harmonic chain. To demonstrate this we compute OTOC on a driven dissipative harmonic chain starting with initial condition $\{y_i(0), \dot{y}_i(0)\}$ taken 
from the position and velocity configurations of the original non-linear DC at a time $\tau > \tilde{\tau}$. In Fig.~\ref{fig:tc_harmonic} we observe good agreement between the OTOC of the original system with that obtained from the effective driven dissipative harmonic system.

More precisely, we compute the OTOC using following two dynamics: (i) Original evolution given in Eq.~\eqref{eqm} corresponding to on-site double well potential $V(x_{i})=(-\frac{x_i^2}{2}+\frac{x_i^4}{4})$ $\forall i$.  (ii) Evolution obtained by performing harmonic approximation of the double well potential for each oscillator around one of the wells in which the oscillator is at some large time~$\tau$, in the original dynamics. 
 If $\{y_i(\tau)\}$ be the positions of the oscillators in the dynamics (i) at time $\tau$, then in the dynamics (ii) we approximate the 
double potential by $\tilde{V}(y_{i}) \approx -\frac{1}{2}+(y_i -\delta_i)^2$ where $\delta_i=1$ if $y_i(\tau)$ falls in the well on the positive side and $-1$ otherwise. The heat-maps corresponding to these two dynamics are shown in Figs.~\ref{fig:tc_harmonic} (a) and (b) from $\tau=600$ to $\tau=800$.  We observe that these two plots resemble quite closely implying that after large time the oscillators enter from transiently chaotic to non-chaotic region where the DC effectively behaves like a driven dissipative harmonic chain.

\begin{figure}[t]
  \centering \includegraphics[width=8 cm]{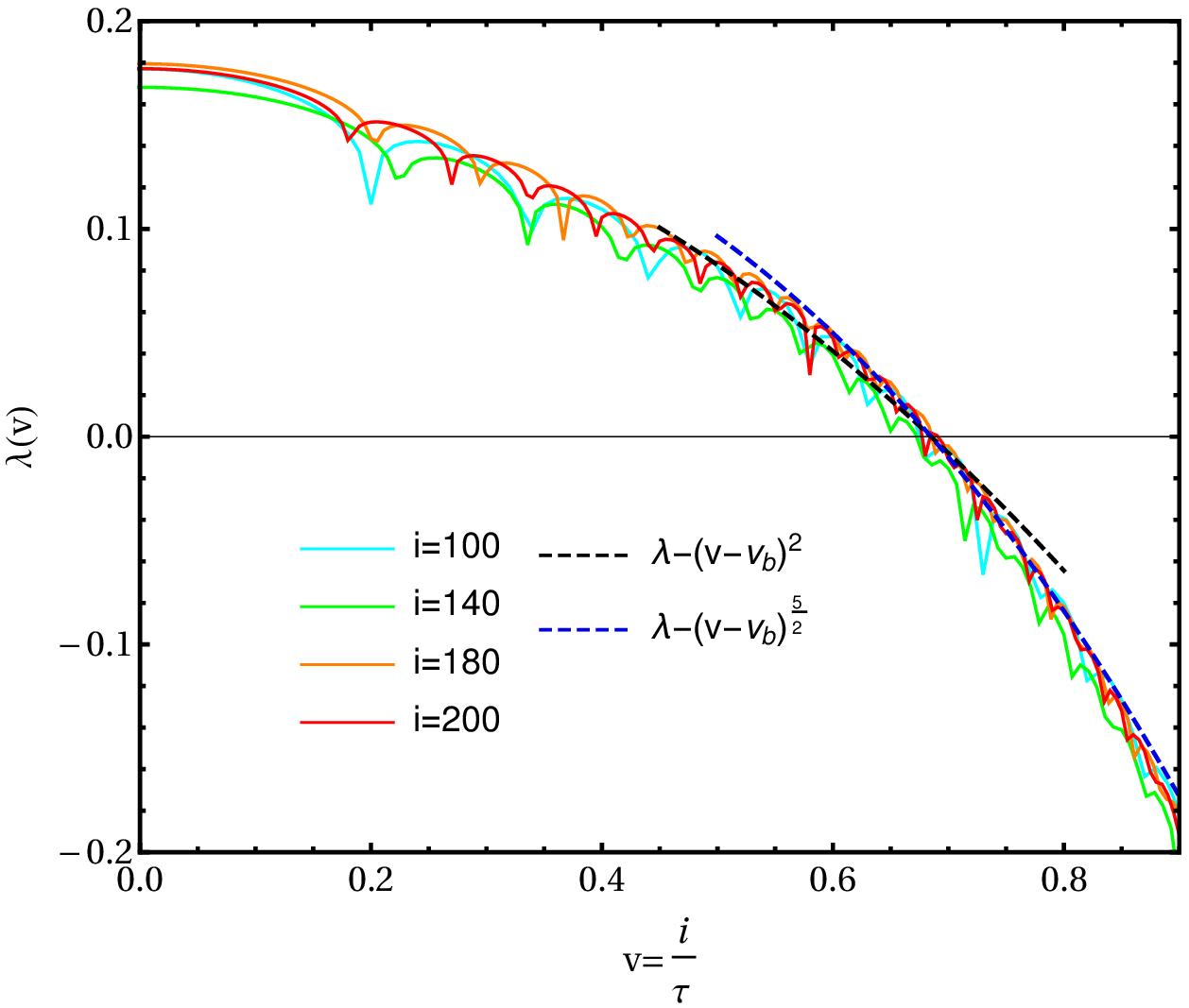}
  \caption{\textit{transient chaos:} $\left\lbrace \bar{f}=0.24,\bar{\gamma}=0.15,\bar{\kappa}=1.0 \right\rbrace$.   
  At comparatively short time ($\tau\approx204$) up to which the light cone exists (marked by the dotted lines in Fig.~\ref{fig:tc_boundary} and Fig.~\ref{fig:tc_ftle}), the DC behaves chaotically and the VDLE falls as $\lambda(v)=\lambda(1-\left(\frac{v}{v_b}\right)^\nu)$ with $\nu=2$ and  $\nu=\frac{5}{2}$ for $v\leqslant v_b$ and $v\geqslant v_b$ respectively. This is similar with the $\lambda(v)$ behavior in sustained chaos regime as observed in  Fig.~\ref{muv-sc-1}.}
\label{fig:tc_vdle_st}
\end{figure}

\begin{figure}[t]
  \centering \includegraphics[width=8 cm]{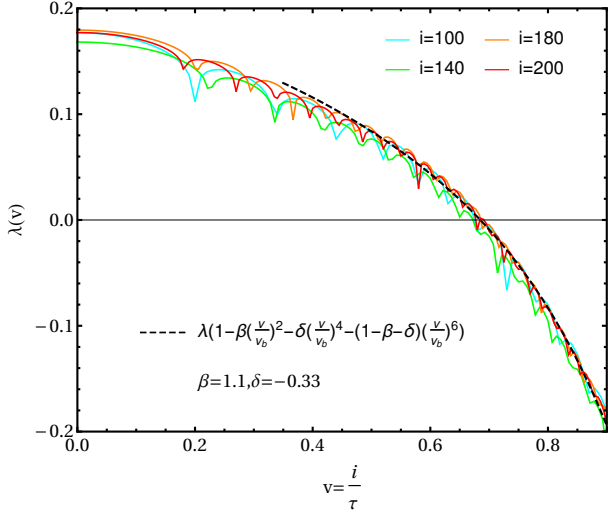}
  \caption{\textit{transient chaos:} $\left\lbrace \bar{f}=0.24,\bar{\gamma}=0.15,\bar{\kappa}=1.0 \right\rbrace$.   
  At comparatively short time ($\tau\approx204$) up to which the light cone exists (marked by the dotted lines in Fig.~\ref{fig:tc_boundary} and Fig.~\ref{fig:tc_ftle}), the DC behaves chaotically and the VDLE falls in the same way $\lambda(v)=\left[1-\beta(v/v_b)^2-\delta(v/v_b)^4-(1-\beta-\delta)(v/v_b)^6\right]$ with $\beta=1.10$ and $\delta=-0.33$ for both inside and outside the light-cone  near $v\approx v_b$. This is similar with the $\lambda(v)$ behavior in sustained chaos regime as observed in  Fig.~\ref{muv-sc-2}.}
\label{fig:tc_vdle_st_1}
\end{figure}

Until now we have observed that in this case, the dynamics of DC crosses over from a chaotic regime to non-chaotic regime through a transient regime as demonstrated in the evolution of FTLE and heat-map plot. 

We now investigate how this crossover gets manifested through VDLE. Following the same procedure as done in the previous two sections, we compute VDLE in the two regimes: $\tau < \tau^*$ and $\tau > \tilde{\tau}$. 



\begin{figure}[t]
  \centering \includegraphics[width=8 cm]{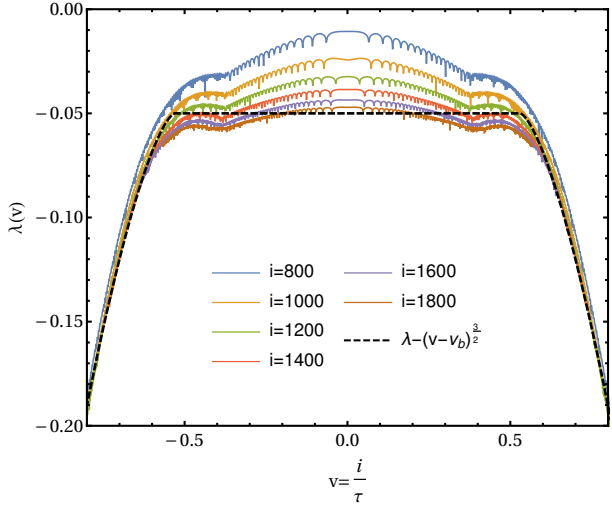}
  \caption{\textit{transient chaos:} $\left\lbrace \bar{f}=0.24,\bar{\gamma}=0.15,\bar{\kappa}=1.0 \right\rbrace$.   At large time ($\tau\geqslant800$), the VDLE is negative meaning the whole DC becomes non-chaotic. Here the VDLE satisfy the relation $(\lambda(v)-\lambda)\sim -(v-v_b)^{3/2}$ for $~v\geqslant v_b$. This is consistent with the $\lambda(v)$ behavior in the non-chaotic regime as observed in  Fig.~\ref{fig:muv-nc-1}.}
\label{fig:tc_vdle_lt}
\end{figure}

In Fig.~\ref{fig:tc_vdle_st} we have plotted $\lambda(v)$ vs $v$ for $\tau\leqslant\tau^*$. For reasons already discussed in section \ref{sec:sc} for the sustained chaos case, we first try to fit  the function $ \lambda \left(1-\left(\frac{v}{v_b}\right)^\nu\right)$ to the VDLE curve in Fig.~\ref{fig:tc_vdle_st}. It is observed that  the data around $v \simeq v_b$ fits well with the following exponents as given below, 
\begin{align}
\label{slopestc}
\lambda(v)=
\begin{cases}
\lambda\left[1-(v/v_b)^2\right],~\text{for}~v\lesssim v_b \\
\lambda\left[1-(v/v_b)^{5/2}\right],~\text{for}~v \gtrsim v_b. \\
\end{cases}
\end{align}
A subsequent search for a single functional form of VDLE that holds for both $v<v_b$ and $v>v_b$ reveals that,
\bea
\lambda(v)&=&\left[1-\beta(v/v_b)^2-\delta(v/v_b)^4-(1-\beta-\delta)(v/v_b)^6\right],\cr
\beta&=&1.10, ~~\delta=-0.33,
\label{eq:vdle_tc_st}
\eea
near $v\approx v_b$ which is presented in Fig.~\ref{fig:tc_vdle_st_1}. On the other hand for $\tau>\widetilde{\tau}$,  we observe in Fig.~\ref{fig:tc_vdle_lt} that the data around $v \simeq v_b$ fits well with the following form 
\begin{align}
\lambda(v)=
\begin{cases}
\lambda,~~~~~~~~~~~~~~~~~\text{for}~v \lesssim v_b \\
\lambda-(v-v_b)^{\frac{3}{2}},~\text{for}~v \gtrsim v_b. \\
\end{cases} \label{eq:muv_tc}
\end{align}
This is expected since the system has made a transit  from chaotic to non-chaotic regime so that the VDLE here in Eq. (\ref{eq:muv_tc}) behaves in the same way as obtained in Eq. (\ref{eq:muv_nc}) for the non-chaotic scenario.

\section{Variation of FTLE and IS with $\bar{f},\bar{\gamma},\bar{\kappa}$}
\label{lvbvsp}
So far, we have chosen parameters such that we are in a particular regime of interest such as chaotic, non-chaotic or transient regimes. In this section we study what happens if we tune parameters so that we go through all the regimes. In particular we vary $\bar{f}$ or $\bar{\gamma}$ or $\bar{\kappa}$ continuously and observe how do the FTLE [$\lambda_i(\tau)$]  or IS [$v_b(\tau,D_\mathrm{th})$] change as we cross from one regime to another. Such studies are important 
in diverse areas such as optimal signal transmissions, secure communications, synchronization in electronic circuits \cite{Peng_1996,Lakshmanan_1996,Blakely_2000,Ivancevic_2007,Musielak_2009}, where a common goal is to gain control over chaotic systems. In this connection, we should mention that a novel chaotic secure communication system has been proposed in Ref.~\cite{Zapateiro_2013} where the encryption system consists of a Duffing oscillator. However, it is also argued \cite{Zapateiro_2013} that, use of only one Duffing oscillator in encryption stage leads to low level of security. So, one might think of considering the coupled Duffing chain as a plausible candidate for increasing the security level of the encrypted messages in those communication systems.   

\begin{figure}[h]
  \centering
  \subfigure[]{\includegraphics[scale=0.39]{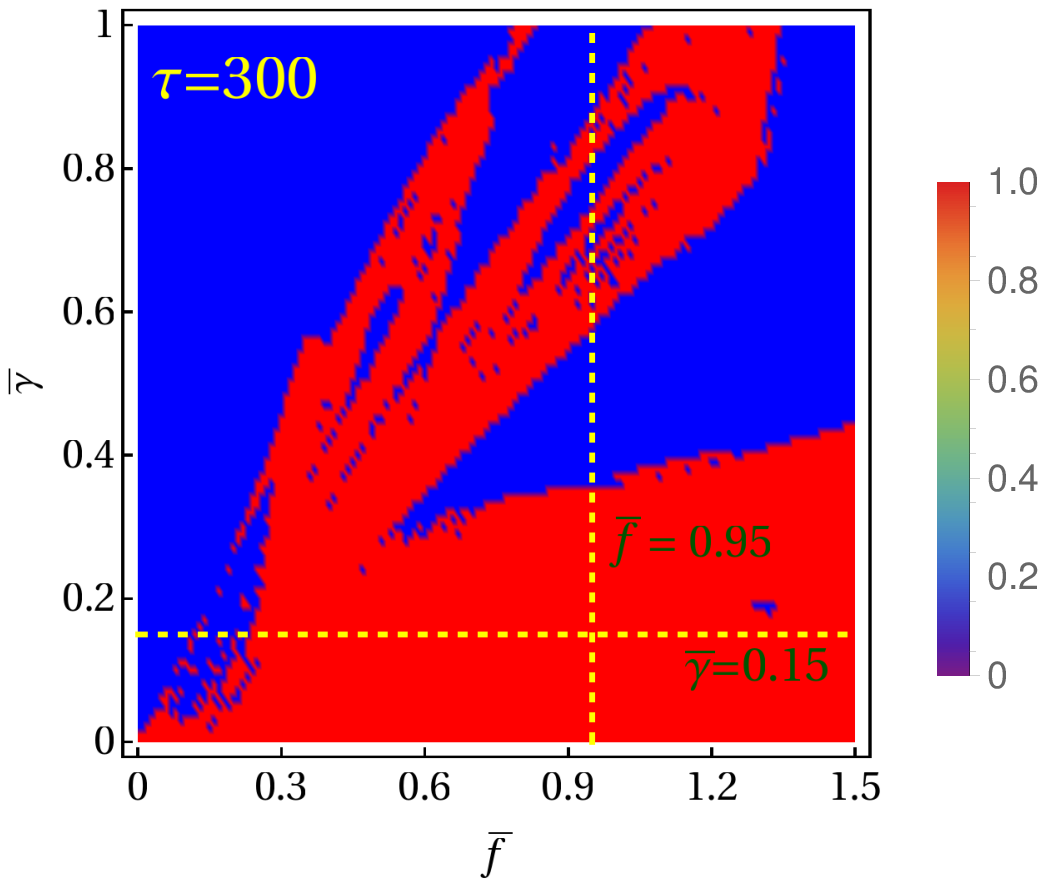}}\hfill
  \subfigure[]{\includegraphics[scale=0.39]{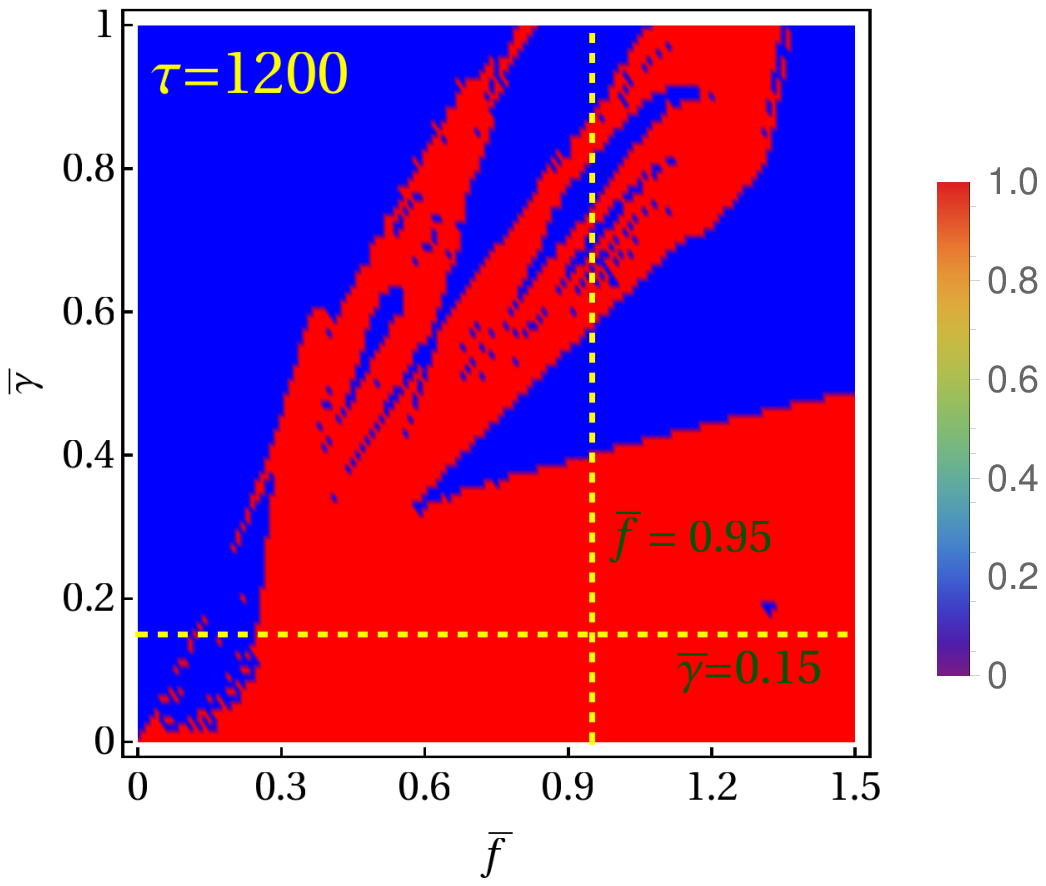}}\\
  \subfigure[]{\includegraphics[scale=0.39]{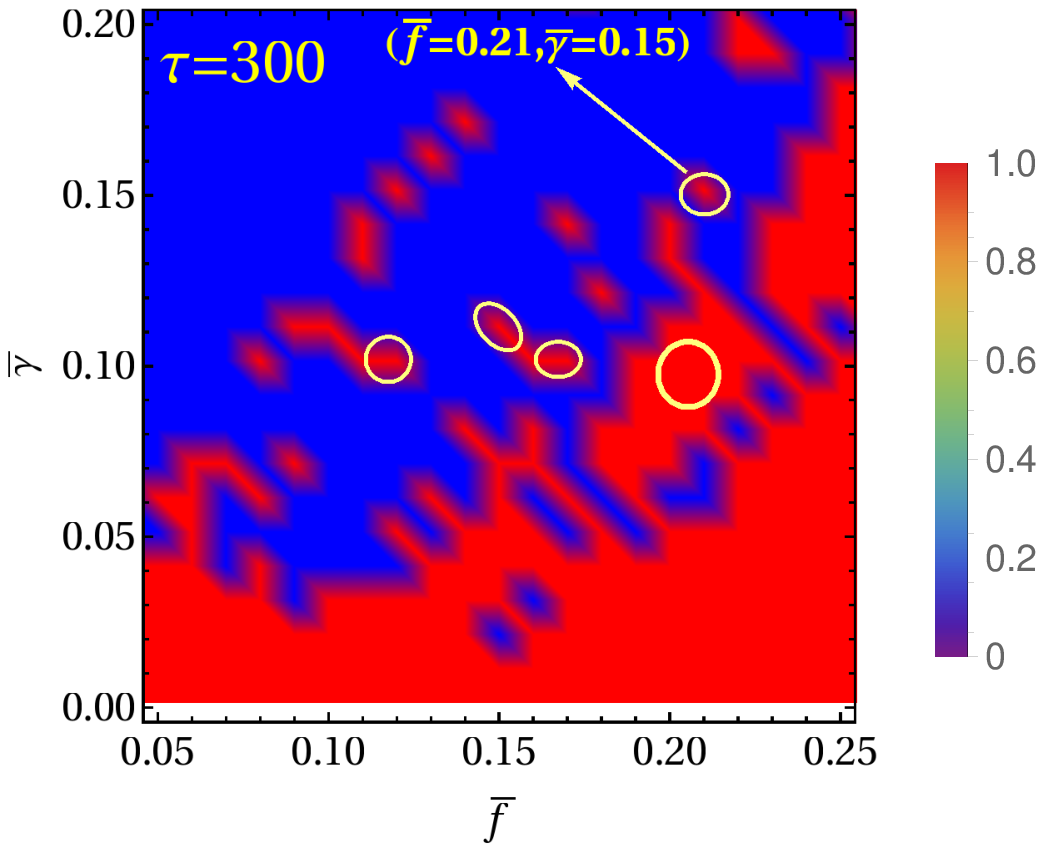}}\hfill
  \subfigure[]{\includegraphics[scale=0.39]{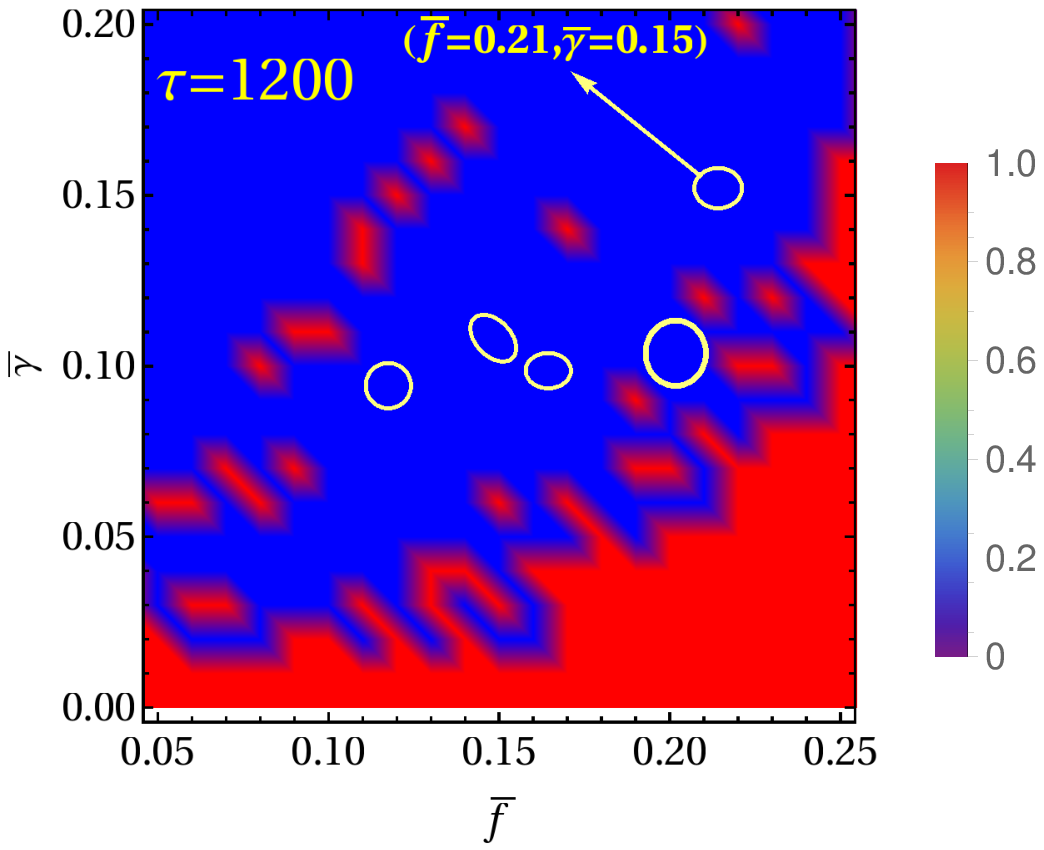}}
  \caption{ $\left\lbrace  \bar{\kappa}=1.0 \right\rbrace$. The heat-maps in  (a) and (b) show the dynamical behavior of DC in $\bar{\gamma}-\bar{f}$ plane as we vary $\bar{f}\in \left[0,1.5\right]$ and $\bar{\gamma}\in \left[0,1\right]$ at time $\tau=300$ (left panel) and $\tau=1200$ (right panel) respectively. The parameter regime with large drive and comparatively small dissipation (e.g. $\bar{f}\in \left[0.5,1.5\right]$) and $\bar{\gamma}\in \left[0,0.3\right]$) is spanned by sustained chaos (red region). Whereas the parameter regime corresponding to low drive and large dissipation (e.g. $\bar{f}\in \left[0,0.2\right]$ and $\bar{\gamma}\in \left[0.3,1\right]$) results in a fully non-chaotic regime (blue region). These fully chaotic and fully non-chaotic regions are separated by regions of highly intermittent dynamical behaviors - for example, in $(\bar{f},\bar{\gamma})\in \left[0.5,1\right]$, we observe a mixture of puddles of chaotic and non-chaotic windows (irregularly occurring red and blue regions). The dynamics of DC along the dashed lines $\bar{\gamma}=0.15$ and $\bar{f}=0.95$ is discussed in detail. To identify the transient chaos regions, in (c) and (d), we plot zoomed in portions of (a) and (b) respectively.  As we shift from the left panel ($\tau=300$) to the right panel($\tau=1200$), we observe the disappearance of the previously existing transient chaos regions. Some of this transient chaos regions are enclosed by yellow rings- which appear as chaotic (red) at $\tau=300$ but becomes non-chaotic (blue) at $\tau=1200$. One such particular parameter set $\left( \bar{f}=0.21,\bar{\gamma}=0.15\right)$ corresponding to transient chaos is pointed out in both the panels and will be discussed more elaborately in the next section.}
\label{heatmapfg}
\end{figure}

For FTLE measurement we choose to study $\lambda_0(\tau)$ and for $v_b(\tau,D_{\text{th}})$ we fix $D_{\text{th}}=1$. Note again that $v_b(\tau,1)$ in different regimes behaves as
\begin{align}
v_b(\tau,1)
\begin{cases}
=0, ~~~~~~~\mathrm{for~large}~\tau~\Rightarrow \mathrm{Non-Chaotic}\\
 \\
=v_b >0, ~\mathrm{for~large}~\tau~\Rightarrow \mathrm{Sustained~chaos}\\ \\
\left.
\begin{array}{c}
>0~~\mathrm{for~small}~ \tau \\
=0 ~~\mathrm{for~large}~ \tau
\end{array}
\right \}
\Rightarrow \mathrm{Transient~chaos}.
\end{cases} \label{IS-decide}
\end{align}
On the other hand, for different choices of parameters we look at the saturation value $\lambda=\lambda_0(\tau)|_{\tau \to \infty}$  to check if the DC belongs to sustained chaos ($\lambda >0$) or non-chaotic ($\lambda < 0$) regimes. In Fig.~\ref{heatmapfg} (a) and (b), we present heat-map plots of $\lambda$ in the $\bar{\gamma}-\bar{f}$ plane at $\tau=300$ and $\tau=1200$ respectively for $\bar{\kappa}=1$. In both the plots, the red regions correspond to sustained chaos regime and the blue regions correspond to the non-chaotic regime.  When comparing between Fig.~\ref{heatmapfg}(a) and (b), a careful observation reveals the disappearance of red regions (and appearance of blue regions accordingly) when going from Fig.~\ref{heatmapfg}(a)($\tau=300$) to Fig.~\ref{heatmapfg} (b) ($\tau=1200$), indicating the existence of transient chaos regimes. To identify these transient chaos regions more appropriately, we zoom in a particular parameter region from Fig.~\ref{heatmapfg}(a) and (b) and plot them in Fig.~\ref{heatmapfg}(c) and (d) respectively. There, we observe that the regions marked by yellow rings are red at earlier time ($\tau=300,$ left panel) whereas they become blue at later time ($\tau=1200,$ right panel), implying that these parameter regions correspond to transient chaos regimes.

In the below sections we discuss, in detail, our numerical results on the variation of  $\lambda_0(\tau)$ and $v_b(\tau,1)$ with respect to one parameter (while keeping the other two fixed) for different times $\tau$. 

\subsection*{Variation with respect to $\bar{f}$}
\begin{figure}[h]
  \centering \includegraphics[width=8 cm]{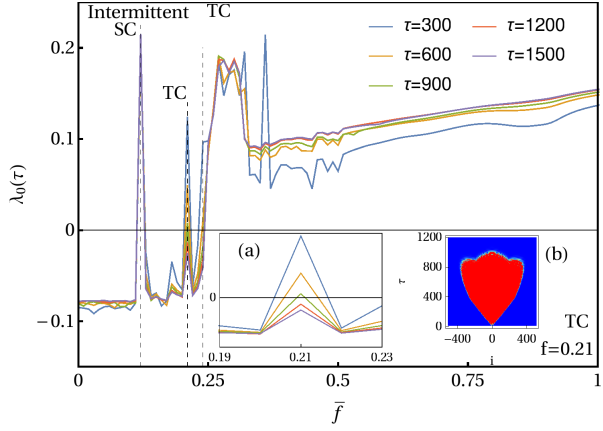}
  \caption{$\left\lbrace\bar{\gamma}=0.15,\bar{\kappa}=1.0\right\rbrace$. The driving amplitude value $\bar{f}^\ast=0.25$ separates the sustained chaos (SC) regime ($\bar{f}>0.25$) from non-chaotic (NC) regime ($0<\bar{f}<0.25$). In $\bar{f}>0.25$ (SC), the FTLE curves at large times saturate to constant $\lambda >0$ whereas the curves in $0<\bar{f}<0.25$ (NC) saturate to  $\lambda <0$. Particularly, deep inside SC, for $0.35<\bar{f}<1,$ $\lambda(\tau)$ is a monotonically increasing function of $\bar{f}$. Inside NC, there are transient chaos points at $\bar{f}=0.21$ (heat-map shown in inset) and $\bar{f}=0.24$ [(heat-map in Fig.~\ref{heatmap}(c)] characterized by crossing of $\lambda(\tau)$ from positive to negative values as time progresses (e.g. for $\bar{f}=0.21$, $\tau=300,600,900$ correspond to $\lambda(\tau)>0$ whereas $\tau=1200,1500$ correspond to $\lambda(\tau)<0$). Also, there exists an intermittent chaotic window at $\bar{f}=0.12$ with $\lambda(\tau)=\lambda>0$ surrounded by NC points all with $\lambda<0$. The measurements are performed every $\Delta\bar{f}=0.01$ on the x-axis.}
\label{fig:lvsf}
\end{figure}

\begin{figure}[h]
  \centering \includegraphics[width=8 cm]{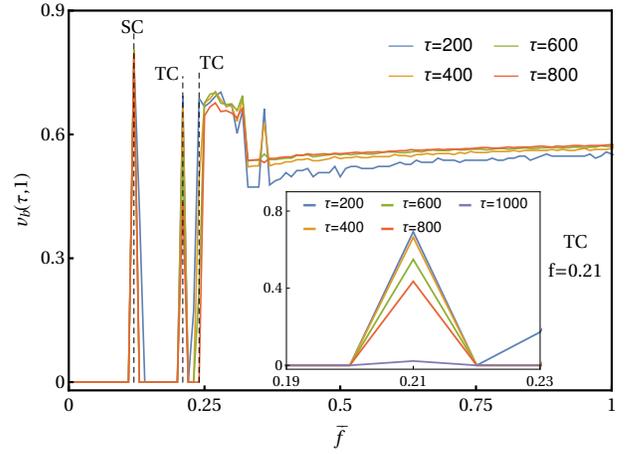}
  \caption{$\left\lbrace\bar{\gamma}=0.15,\bar{\kappa}=1.0\right\rbrace, D_{\mathrm{th}}=1$. The IS characterizes NC regime ($0<\bar{f}<0.25$) with $v_b(\tau)=0$ whereas the IS curves at different times saturate to constant $v_b(\tau)=v_b>0$ in SC regime ($\bar{f}>0.25$). Inside the SC regime, for $0.4<\bar{f}<1.0$, IS is a monotonically increasing but much slowly varying function of $\bar{f}$ in comparison to FTLE. Inside the NC regime ($0<\bar{f}<0.25$), the transient chaos points $\bar{f}=0.21$ (heat-map shown in inset of Fig.~\ref{fig:lvsf}) and $\bar{f}=0.24$ [heat-map in Fig.~\ref{heatmap}(c)] are characterized by a decreasing $v_b(\tau)$ towards zero as time progresses.The intermittent chaotic window at  $\bar{f}=0.12$ (heat-map in inset) has $v_b(\tau)=v_b>0$ inside the otherwise NC regime with all surrounding points with $v_b(\tau)=0$. }
\label{fig:vbvsf}
\end{figure}

In Fig.~\ref{fig:lvsf} and  Fig.~\ref{fig:vbvsf} we plot the variation of $\lambda_0(\tau)$ and  $v_b(\tau,1)$ with respect to $\bar{f}$, respectively, for $\bar{\gamma}=0.15,~\bar{\kappa}=1.0$ and for different values of $\tau$. In both the plots we observe that with  $\bar{f}$, increasing from $0$ to $1$,  the system crosses over from non-chaotic to sustained chaos regime through an intermediate transient regime [$\bar{f} \sim (0.21 - 0.24)$]. However, deep inside the non-chaotic regime we observe some intermittent window of sustained chaos ($\bar{f} = 0.12$).

Although both FTLE and IS are time dependent entities in general, the convergence of the curves at large times clearly indicate the parameter regions giving rise to sustained chaos  regime or non-chaotic regime  as can be observed in Fig.~\ref{fig:lvsf} and Fig.~\ref{fig:vbvsf}.  In particular we observe in  Fig.~\ref{fig:lvsf} that $\bar{f}^\ast=0.25$ separates sustained chaos ($\bar{f}>0.25$) regime with $\lambda_0(\tau\rightarrow\infty)=\lambda>0$ and non-chaotic ($0<\bar{f}<0.25$) with $\lambda_0(\tau\rightarrow\infty)=\lambda<0$. The same sustained chaos and non-chaotic regimes can be alternatively identified from Fig.~\ref{fig:vbvsf} with $v_b(\tau,1)=v_b>0$ and $v_b(\tau,1)=v_b=0$ respectively.

As mentioned earlier, inside the non-chaotic region $0<\bar{f}<\bar{f}^\ast,$ we interestingly observe intermittent points of sustained chaos (e.g. at $\bar{f}=0.12$) and transient chaos (e.g. at $\bar{f}=0.21$ and $\bar{f}=0.24$). The appearance of transient chaos at $\bar{f}=0.24$ has already been discussed in section \ref{sec:tc}. Here we focus on the transient chaos appearing at $\bar{f}=0.21$ and demonstrate how one can identify this feature from the $\lambda(\tau)$ {\it vs.} $\bar{f}$ and $v_b(\tau)$ {\it vs.} $\bar{f}$ plots for different $\tau$. In the inset (a) of Fig.~\ref{fig:lvsf} we zoom the behavior of $\lambda_0(\tau)$ near $\bar{f}=0.21$ where  we note that at smaller $\tau,$ the FTLE $\lambda_0(\tau)>0$ suggesting the dynamics could be chaotic. But with increasing $\tau$, we observe that the value of FTLE at $\bar{f}=0.21$ decreases and finally at large $\tau$ it saturates to a value $\lambda<0$. This indicates that the dynamics for $\bar{f}=0.21$ is actually transient which crosses over from sustained to non-chaotic regime as time progresses. For reference, a heat map plot of the OTOC at $\bar{f}=0.21$ is also shown in the inset (b) of Fig.~\ref{fig:lvsf}. Alternatively, the same feature at this value of $\bar{f}$ can be observed from the $v_b(\tau,1)$ {\it vs.} $\bar{f}$ plots for different values of $\tau$ in Fig.~\ref{fig:vbvsf}, where the crossover is demonstrated (see the inset of  Fig.~\ref{fig:vbvsf}) by the decrease of IS to zero with increasing $\tau$ as mentioned in Eq.~\eqref{IS-decide}.

Deep inside the SC regime, for $\bar{f}\in\left[0.4,1\right]$, it is observed that FTLE grows linearly with $\bar{f}$. In connection to this observation, it is worth mentioning that there has been a recent conjecture $\lambda\propto \sqrt{T}$ in classical chaotic Hamiltonian systems where $T$ is the temperature \cite{Kumar_2019}. Our observation $\lambda\propto \bar{f}$ in  Fig.~\ref{fig:lvsf} is consistent with this conjecture as the energy scale of each oscillator in the SC regime is $\sim \bar{f}^2$ which can be considered as effective temperature in our  driven dissipative system. On the other hand, as observed in Fig.~\ref{fig:vbvsf}, $v_b(\tau,1)$ almost remain constant as we vary $\bar{f}$ inside the sustained chaos regime. This indicates that the driving amplitude ($\bar{f}$) has more impact on the FTLE than on IS.

\subsection*{Variation with respect to $\bar{\gamma}$}
\begin{figure}[h]
  \centering \includegraphics[width=8 cm]{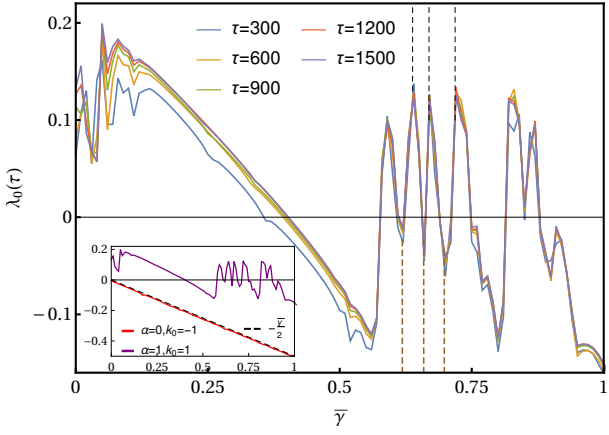}
  \caption{$\left\lbrace\bar{f}=0.95,\bar{\kappa}=1.0\right\rbrace$. The dissipation value $\bar{g}^\ast=0.4$ separates the sustained chaos (SC) regime ($0<\bar{\gamma}<0.4$) from non-chaotic (NC) regime ($\bar{\gamma}>0.4$). In $0.05<\bar{\gamma}<0.55,$ FTLE is a monotonically decreasing function of $\bar{\gamma}$. The FTLE curves at different $\tau$ saturate to $\lambda(\tau)=\lambda>0$ in SC and to $\lambda(\tau)=\lambda<0$ in NC regime. The parameter regime $0.55<\bar{\gamma}<0.85$ (inside NC) exhibits a highly intermittent behavior with several chaotic windows. The computation is done with the resolution $\Delta\bar{\gamma}=0.01$ along the $x$-axis.}
\label{fig:lvsg}
\end{figure}

\begin{figure}[h]
  \centering \includegraphics[width=8 cm]{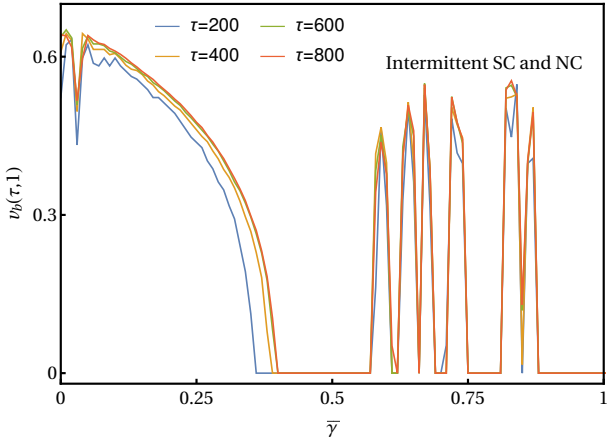}
  \caption{$\left\lbrace\bar{f}=0.95,\bar{\kappa}=1.0\right\rbrace, D_{\mathrm{th}}=1$. In SC regime ($0<\bar{\gamma}<0.4$), $v_b(\tau)$ curves for different $\tau$ saturate to $v_b(\tau)=v_b>0$ and behave in a monotonically decreasing way as $\bar{\gamma}$ is increased. Inside the NC regime ($0.4<\bar{\gamma}<1$), the intermittent chaotic windows have $v_b(\tau)=v_b>0$ surrounded by all NC points with $v_b(\tau)=v_b=0$.}
\label{fig:vbvsg}
\end{figure}

In this section we study the variation of $\lambda_0(\tau)$ and $v_b(\tau,1)$ with respect to $\bar{\gamma}$ for $\bar{f}=0.95$ and $\bar{\kappa}=1.0$ at different values of $\tau$. In  Fig.~\ref{fig:lvsg} and  Fig.~\ref{fig:vbvsg} we plot the variation of $\lambda_0(\tau)$ and $v_b(\tau,1)$, respectively, over the range $\bar{\gamma} \in \left[0,1\right]$.  As noticed earlier, the curves at large $\tau$ converge in both the figures. We see that the value $\bar{\gamma}^\ast=0.4$ marks the transition from SC regime [$\bar{\gamma}\in(0,0.4)$] to NC regime [$\bar{\gamma}\in(0.4,1.0)$]. It seems that FTLE in Fig.~\ref{fig:lvsg} decreases approximately linearly with increasing $\bar \gamma$ in the regime $0.05<\bar{\gamma}<0.55$. 
This sustained chaos regime is identified by a monotonic but non-linear decrease of $v_b(\tau,1)$ with increasing $\bar{\gamma}$ in Fig.~\ref{fig:vbvsg} . It is interesting to observe that this monotonic decrease in FTLE and IS is followed by a highly intermittent behavior as we further increase the dissipation [$\bar{\gamma}\in(0.55,1)$]. In particular, we observe a mixture of chaotic and non-chaotic windows in this parameter regime from both Fig.~\ref{fig:lvsg} and  Fig.~\ref{fig:vbvsg}.

To understand the above mentioned approximately linear decrease of the FTLE for $0.05<\bar{\gamma}<0.55$, we  look at how does FTLE  $\lambda_0(\tau)$ vary with increasing $\bar \gamma$ for a driven-dissipative harmonic chain. For this case it is possible to compute FTLE analytically (see Appendix.~\ref{AI}) 
and we find $\lambda_0(\tau)$ decays linearly as $\lambda_0(\tau)=-\frac{\bar{\gamma}}{2}$. In the inset of Fig.~\ref{fig:lvsg}, a comparison between the  FTLE of the harmonic chain and of the Duffing chain is provided. We observe that the FTLE in the anharmonic case decays with $\bar \gamma$ although the dynamics at small $\bar \gamma$ is chaotic in contrast to the harmonic case for which the dynamics is always non-chaotic as expected. However, upon increasing $\bar \gamma$ further the FTLE goes beyond zero and becomes negative till $\bar \gamma =0.56$ after which the behavior with respect to $\bar \gamma$ becomes irregular with chaotic and non-chaotic regimes appearing apparently abruptly.

%
\begin{figure}[h]
  \centering \includegraphics[width=8 cm]{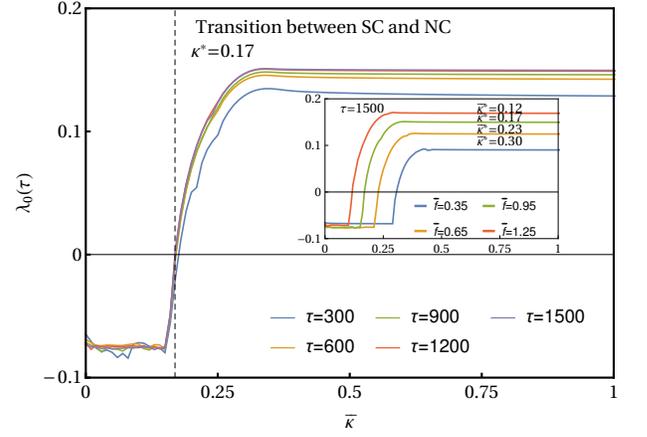}
  \caption{$\left\lbrace\bar{f}=0.95,\bar{\gamma}=0.15\right\rbrace$ for the main figure. For $\kappa=0,$ $\lambda_0(\tau)=\lambda<0$ implies the uncoupled DC is non-chaotic. As the coupling is turned on and increased, the DC becomes chaotic at $\bar{\kappa}^\ast=0.17$. In non-chaotic ($0<\bar{\kappa}<0.17$) and sustained chaos ($\bar{\kappa}>0.17$) regime the FTLE curves at different $\tau$ saturate to $\lambda_0(\tau\rightarrow\infty)=\lambda<0$ and $\lambda_0(\tau\rightarrow\infty)=\lambda>0$ respectively. It is also observed that deep inside the sustained chaos (SC) and non-chaotic (NC) regimes,   $\lambda_0(\tau\rightarrow\infty)$ is a very slowly varying function of $\bar{\kappa}$. In the inset ($\bar{\gamma}=0.15, \tau=1500$), we observe that with increasing driving amplitude ($\bar{f}$), the minimum coupling ($\bar{\kappa}^\ast$) required to make the DC chaotic decreases.}
\label{fig:lvsk}
\end{figure}
\begin{figure}[h]
  \centering \includegraphics[width=8 cm]{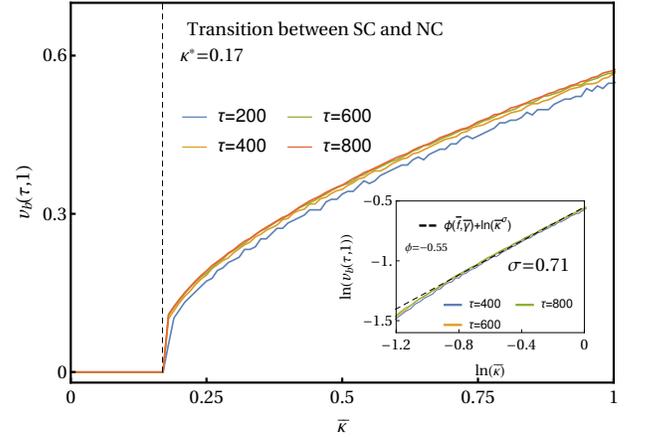}
  \caption{$\left\lbrace\bar{f}=0.95,\bar{\gamma}=0.15\right\rbrace$. In the sustained chaos (SC) ($0.17<\bar{\kappa}<1$) regime, the IS curves at different $\tau$ converge to $v_b(\tau,1)=v_b>0$ whereas we see $v_b(\tau,1)=v_b=0$ in the non-chaotic (NC) ($0<\bar{\kappa}<0.17$) regime. Inside the SC, for comparatively large $\kappa$ ($\kappa>0.4$), the IS varies with $\bar{\kappa}$ as $v_b \sim \kappa^{\sigma}$ with $\sigma=0.71$. That $\sigma=0.71$ is shown in the inset by plotting the main figure in log-log scale.}
\label{fig:vbvsk}
\end{figure}

\subsection*{Variation with respect to  $\bar{\kappa}$}
Here we would like to discuss the effect of the coupling ($\bar{\kappa}$) on $\lambda_0(\tau)$ and $v_b(\tau,1)$ at different $\tau$ while the other parameters are fixed to $\bar{f}=0.95,\bar{\gamma}=0.15$.  From Fig.~\ref{fig:lvsk} and Fig.~\ref{fig:vbvsk}, we note that when $\bar{\kappa}=0$ i.e. for uncoupled Duffing oscillators, the system is non-chaotic with $\lambda_0(\tau)=\lambda<0$ and $v_b(\tau,1)=0$ respectively.

In Fig.~\ref{fig:lvsk}, it is very interesting to observe, as we turn on the coupling $\bar{\kappa}$, that near $\bar{\kappa}^\ast=0.17,$ the DC transits from non-chaotic ($\lambda_0(\tau)=\lambda<0$) to sustained chaos regime ($\lambda_0(\tau)=\lambda>0$). 
Equivalently, this feature manifests itself as a transition from $v_b(\tau,1)=v_b<0$ to $v_b(\tau,1)=v_b>0$ in Fig.~\ref{fig:vbvsk}. So, this behavior of the DC indicates that coupling alone can initiate chaotic behavior in spatially extended systems.

A natural question one might ask is, how does the minimum coupling strength $\bar{\kappa}^\ast$,  required to make the system chaotic, change as we vary the driving amplitude $\bar{f}$ ? This is an  important question given the possibility of tuneability of various parameters such as coupling and driving.  To answer this, we present the behavior of $\lambda_0(\tau)$ versus $\bar{\kappa}$ for different values of $\bar{f}$ at large time in the inset of Fig.~\ref{fig:lvsk}. We observe that as we increase the driving amplitude $\bar{f},$ $\bar{\kappa}^\ast$ decreases which implies at higher $\bar{f}$ comparatively lower coupling is sufficient to turn on the chaos in the DC. However, for $\bar{\kappa}\in (0.35,1)$ i.e. deep inside the chaotic regime, and $\bar{\kappa}\in (0,0.15)$, i.e., deep inside the non-chaotic regime we observe from Fig.~\ref{fig:lvsk} that the FTLE is almost independent of $\bar{\kappa}$ as manifested by plateau regions on the right and left  of the $\bar{\kappa}^*$. 

On the other hand, in the sustained chaos regime corresponding to the range $0.17<\bar{\kappa}<1.0$, from Fig.~\ref{fig:vbvsk} we see that $v_b(\tau,1)$ increases significantly with $\bar{\kappa}$ as a power law. In this regard, one may note that even in the absence of non-linearity (driven-dissipative harmonic chain), $v_b$ behaves as a power-law $v_b \sim \sqrt{\kappa}$ (see Appendix.~\ref{AI}).
In the presence of non-linearity, as shown in the inset of Fig.~\ref{fig:vbvsk}, it is interesting to observe that deep inside the sustained chaos regime, corresponding to $\bar{\kappa}\in (0.4,1)$,  at large time, the IS follows the functional form $v_b \sim \kappa^{\sigma}$ with $\sigma=0.71$.  The fact that $\sigma$ is different from 0.5 is a fingerprint of non-linearity which explicitly includes the effect of drive strength in contrast to the harmonic chain case. 

\section{Conclusions and Outlook}
In this paper, we have studied the dynamics of a driven dissipative chain of coupled Duffing oscillators. Interestingly, depending on the choice of the system parameters (namely the driving amplitude, driving frequency, dissipation, nonlinearity and coupling strength), the Duffing chain is observed to exhibit rich dynamical behavior with three different dynamical regimes -  (i) sustained chaos, (ii) non-chaotic regime and (iii) transient chaos. Although the existence of these dynamical regimes were known~\cite{Umberger_1989}, powerful diagnostics to investigate these rich regimes have been missing. 

\begin{table*}[ht]
  \begin{center}
    \begin{tabular}{|p{0.12\textwidth}|p{0.23\textwidth}|p{0.28\textwidth}|p{0.33\textwidth}|}
    \hline
     \,  & \textbf{Sustained chaos} & \textbf{Non-chaotic regime}\hspace{1 cm} & \textbf{Transient  chaos}\hspace{1.5 cm} \\
      \hline
      \textbf{OTOC} & exponential growth and & exponential decay and  ballistic\hspace{0.6 cm} &  Dynamical crossover: exponential growth \\ 
      $[D(i,\tau)]$ & ballistic spread &  spread in $D_{\mathrm{th}}\rightarrow0$ limit\hspace{1 cm} & and ballistic spread$\longrightarrow$ non-growing and  non-spreading OTOC \\
      \, & \, & \, & \hspace{1.5 cm} \\
      {\it Heat-map} & {\it Light-cone with sharp} & {\it Light-cone with non-sharp} & {\it Complex geometric shapes with initial}\hspace{0.5 cm} \\
      {\it structure} & {\it boundaries}\;\; [Fig.~\ref{heatmap} (a)] & {\it  boundaries}\;\; [Fig.~\ref{heatmap} (b)] & {\it light-cone formation}\;\; [Fig.~\ref{heatmap} (c,d)] \\
      \hline
      \textbf{FTLE} & $\lambda_i(\tau\approx0)< 0$ & $\lambda_i(\tau\approx0)< 0$ \hspace{1.5 cm} & $\lambda_i(\tau\approx0)< 0$ \hspace{2 cm} \\
      $[\lambda_i(\tau)]$ & $\lambda_i(\tau\rightarrow\infty)=\lambda>0$ & $\lambda_i(\tau\rightarrow\infty)=\lambda<0$ \hspace{1 cm} & $\lambda_i(\tau)$ crosses from $>0$ to $<0$ at finite $\tau$\\
      \, & \, & \, & and $\lambda_i(\tau\rightarrow\infty)=\lambda<0$\hspace{1.0 cm} \\
      {\it  Number of} & $n_c= 1$ (Fig.~\ref{fig:ftle_sc}) & $n_c = 0$ (Fig.~\ref{fig:lvst_nc})\hspace{2.0 cm} &  $n_c=2$ (Fig.~\ref{fig:tc_ftle})\hspace{2.0 cm} \\
      {\it crossings with  $\lambda_i(\tau)=0$ line} & \, & \, & \, \\
      \hspace{0.7 cm}($n_c$) & \, & \, & \, \\
      \hline
      \textbf{IS} & saturates to constant $v_b>0$, & well-defined only in $D_{\mathrm{th}}\rightarrow0$ limit & \hspace{1.0 cm} $>0$ for small $\tau$\hspace{2 cm} \\
      $[v_b(\tau,D_{\mathrm{th}})]$ & \hspace{0.5 cm} independent of $D_{\mathrm{th}}$ & \hspace{1.0 cm} (inset Fig.~\ref{fig:nc_vb}) & \hspace{1.0 cm} $=0$ for large $\tau$  \\
      \, & \hspace{1.0 cm}(inset Fig.~\ref{fig:nc_vb}) & \, & \hspace{1.2 cm} (inset Fig.~\ref{fig:tc_boundary}) \\
      \, & \, & \, & \, \\
      $D_{\mathrm{th}}=1$ & $v_b(\tau\,\, \mathrm{large},1)=v_b >0$ & \hspace{0.7 cm} $v_b(\tau\,\, \mathrm{large},1)=0$\hspace{1.5 cm} & \hspace{0.8 cm} $v_b(\tau\,\, \mathrm{small},1)>0$  \\
      \, & [Fig.~\ref{fig:vbvsf}, $\bar{f}\in (0.25,1.0)$] & \hspace{0.5 cm} [Fig.~\ref{fig:vbvsf}, $\bar{f}\in (0,0.25)$]\hspace{1.0 cm} & \hspace{0.5 cm} $\&$ $v_b(\tau\,\, \mathrm{large},1)=0$\,\, (inset Fig.~\ref{fig:vbvsf}) \\
      \hline
      \textbf{VDLE} [$\lambda(v)$] & \, & \, & \, \\
      \hspace{0.5 cm} $v\lesssim v_b$ & $\lambda\left[1-\left(\frac{v}{v_b}\right)^2\right]$ (Fig.~\ref{muv-sc-1})  & $\lambda(v)-\lambda\approx0$ (Fig.~\ref{fig:muv-nc-1}) & \hspace{0.7 cm} $\lambda\left[1-\left(\frac{v}{v_b}\right)^2\right]$, \hspace{0.3 cm}small $\tau$ (Fig.~\ref{fig:tc_vdle_st}) \\
      \, & \, & \, & \hspace{0.7 cm} $\lambda(v)-\lambda\approx0$,  \hspace{0.4 cm} large $\tau$ (Fig.~\ref{fig:tc_vdle_lt}) \\
      \hspace{0.5 cm} $v\gtrsim v_b$ & $\lambda\left[1-\left(\frac{v}{v_b}\right)^{\frac{5}{2}}\right]$ (Fig.~\ref{muv-sc-1})  & $(\lambda(v)-\lambda)\sim -(v-v_b)^{\frac{3}{2}}$ (Fig.~\ref{fig:muv-nc-1}) & \hspace{0.7 cm} $\lambda\left[1-\left(\frac{v}{v_b}\right)^{\frac{5}{2}}\right]$, \hspace{0.3 cm} small $\tau$ (Fig.~\ref{fig:tc_vdle_st}) \\
      \, & \, & \, & $(\lambda(v)-\lambda)\sim -(v-v_b)^{\frac{3}{2}}$, large $\tau$ (Fig.~\ref{fig:tc_vdle_lt}) \\
      \hline
    \end{tabular}
     \caption{Characterization of the dynamical regimes of a driven dissipative Duffing chain. The corresponding figures are mentioned alongside.}
     \label{tab:table}
  \end{center}
\end{table*}

We have thoroughly investigated these dynamical regimes by introducing out-of-time-ordered correlator (OTOC) as a promising tool,  which serves as a measure of both the spatial spread and temporal growth (or decay) of an initially localized infinitesimal perturbation. We have observed that spatio-temporal heat-maps  of OTOC (Fig.~\ref{heatmap}) clearly demonstrates the existence of the different dynamical regimes. While the OTOC grows exponentially in the sustained chaos regime, it decays in the non-chaotic regime. In the transient regime we have found that the OTOC at small times looks like the sustained chaos pattern but at large time it crosses over to the non-chaotic regime where it 
ceases both to grow exponentially and spread ballistically. 

In order to quantify separately the spatial spread and and temporal growth (or decay) of the OTOC, we have looked at the instantaneous speed [IS, Eq. (\ref{IS})] and the finite time Lyapunov exponent [FTLE, Eq. (\ref{ftle})], defined directly from the OTOC. Equivalently, to characterize the temporal growth (or decay) of perturbation in a frame moving with a velocity ($v$)  with respect to the initially perturbed oscillator, we have used the velocity dependent Lyapunov exponent [VDLE, $\lambda(v)$] as a spatio-temporal measure of the dynamics, defined directly from OTOC in Eq. (\ref{vdle}). Through extensive numerical simulation and theoretical arguments, we show that these quantities characterise the above mentioned regimes very well. 

We have shown that for all three regimes the FTLE starts from a value $<0$ initially and finally saturates to a non-zero value. For sustained chaos we find that the FTLE saturates to a $\lambda>0$, thus crossing the $\lambda=0$ value only once, whereas for the non chaotic case it never becomes positive and saturates, at large time to a negative value. In the case of transient chaos regime, the FTLE for some oscillators, starting from a negative value increases to a positive value and at a certain time, the FTLEs of all these oscillators start decreasing simultaneously  and finally, at large time saturates to a negative value. Thus in this regime FTLE crosses the $\lambda=0$ line twice. 
We also have shown that IS also provides a good diagnostic for the detection of the three regimes. In particular, we have found that the VDLE in the three regimes behaves distinctly in the sustained regime and in the non-chaotic regime whereas in the transient regime, as for the other two diagnostics, show behaviors similar to sustained chaos at small times and behaviors similar to non-chaotic regimes at late times. All these features are summarized in TABLE \ref{tab:table}.

We have also studied the behavior of FTLE and IS when the driving amplitude ($\bar{f}$), dissipation ($\bar{\gamma}$) and coupling strength ($\bar{\kappa}$) are changed separately.  Such studies are particularly important in 
context of gaining control and tune-ability over chaotic systems. In all three cases, we find that typically the sustained chaos regime and the non-chaotic regimes are separated by a transient chaos regime with intermittent sustained chaos points appearing inside the non-chaotic regime. When $\bar{f}$ is increased from small value the DC undergoes a transition from non-chaotic to sustained chaos regime (see Fig.~\ref{fig:lvsf} and Fig.~\ref{fig:vbvsf}). Deep inside the sustained chaos regime, interestingly, the saturated FTLE increases linearly with $\bar{f}$ (Fig.~\ref{fig:lvsf}).  Similar observations are made from the variation of IS with changing $\bar{f}$ as well, only difference being that the IS does not change much with the increasing drive deep inside the chaotic region.  On the other hand,  we have observed a monotonic and linear decrease in FTLE (Fig.~\ref{fig:lvsg}) and a non-linear decrease in IS (Fig.~\ref{fig:vbvsg}) with increasing dissipation in the sustained chaos region.  This is followed by a highly intermittent mixture of chaotic and periodic windows as one further increases the dissipation. In the case of variation with respect to coupling strength ($\kappa$), the most important observation (see Fig.~\ref{fig:lvsk}) that we made is a follows:  by turning the harmonic coupling only, it is possible to make the dynamics of the DC transit from non-chaotic to chaotic regime and this happens at a critical strength $\kappa^*$ which decreases with increasing driving amplitude. We observe that IS varies with $\kappa$ as $v_b \sim \kappa^{\sigma}$ (see Fig.~\ref{fig:vbvsk}). Interestingly, $\sigma$ for DC is found to be different from $\frac{1}{2}$  obtained for a driven dissipative harmonic chain.

Our work can be explored further in several directions. Since most of our findings rely on extensive numerical simulation,  it would be very interesting to explore possible analytical means of describing the numerical results obtained in this work. In particular we feel it would be possible to develop a perturbation theory for capturing the non-chaotic to chaotic crossover. Another interesting direction to explore is the sensitivity to initial conditions \cite{Lai_1994}. In the present paper, we have dealt with a fixed initial condition. One could investigate the sensitivity of the dynamical properties to different sets of initial conditions. 
A crucial direction is to investigate the effect of adding a stochastic noise \cite{Wei_1997} on the dynamical behavior of the driven-dissipative Duffing chain. To study the generality of the results obtained here, one can consider different systems like self-sustained chain of oscillators e.g. coupled Van der Pol oscillators or coupled Van der Pol-Duffing oscillators \cite{Wei_2011} to analyze the intricate interplay between the self-sustained characteristic with external drive, dissipation and coupling. Having a handle on the classical driven dissipative system, it is a fascinating and a challenging task to study the quantum version of these models \cite{Roy_2001,Pokharel_2018}.

\section*{Acknowledgements}
We acknowledge support of the Department of Atomic Energy, Government of India, under project no.12-R\&D-TFR-5.10-1100.
We also acknowledge Subhro Bhattacharjee, Samriddhi Sankar Ray, Abhishek Dhar, David Huse, Deepak Dhar and Urna Basu for useful discussions. MK would like to acknowledge support from the project 6004-1 of the Indo-French Centre for the Promotion of Advanced Research (IFCPAR), the Ramanujan  Fellowship  SB/S2/RJN-114/2016  and  the SERB Early Career Research Award ECR/2018/002085 from the Science and Engineering Research Board,  Department of Science and Technology, Government of India.   AK  would  like  to  acknowledge  support  from  the project 5604-2 of the Indo-French Centre for the Promotion of Advanced Research (IFCPAR) and the the SERB Early  Career  Research  Award  ECR/2017/000634  from the Science and Engineering Research Board,  Department of Science and Technology, Government of India. The numerical calculations were done on the cluster  {\it Tetris} at the ICTS-TIFR.

\appendix 
\section{VDLE for driven dissipative linear harmonic chain (HC)}
\label{AI}
In section \ref{sec:nc} we have discussed that, in the non-chaotic regime, the dynamics of the DC essentially acts as a driven dissipative linear harmonic chain. A brief calculation has been demonstrated there for the corresponding behavior of the VDLE in Eq. (\ref{eq:muv_nc}). Here we are going to present a rigorous derivation for the results in Eq. (\ref{eq:muv_nc}) starting from the evolution equation of perturbations [Eq. (\ref{hc_1})] given below, 
\be
 \frac{d^2 \delta y_i}{d\tau^2} =-k_0 \delta y_i - \bar{\gamma} \frac{d \delta y_i}{d\tau} + \bar{\kappa} (\delta y_{i-1}+\delta y_{i+1}-2 \delta y_i),
 \label{hc}
 \ee
 where $1\leqslant i \leqslant N$.  We  consider the same initial conditions as in Eq. (\ref{init}) i.e. $\delta y_i=\epsilon \delta_{i,\frac{N+1}{2}}, \frac{d \delta y_i}{d\tau}=0 $ for $i=1,2 \dots N$. Eq. (\ref{hc}) can be represented in the following matrix form
 \be
 \frac{d^2 \delta \bold{Y}}{d\tau^2}=M \delta \bold{Y} -\bar{\gamma} \frac{d \delta \bold{Y}}{d\tau}, \label{hc1}
 \ee
 where $\delta \bold{Y} =\left(\delta y_1 \dots \delta y_i \dots \delta y_N\right)^{T}$ and the matrix $M$ is the following $N \times N$ matrix
 \be
\hspace*{-0.5 cm} M=\left(
 \begin{array}{ccccccc}
  -2\bar{\kappa} -k_0 & \bar{\kappa} & 0 & 0 & .. & 0 & \bar{\kappa} \\
  \bar{\kappa} & -2\bar{\kappa}-k_0 & \bar{\kappa} & 0 & .. & 0 & 0 \\
  0 & \bar{\kappa} & -2\bar{\kappa}-k_0 & \bar{\kappa} & .. & 0 & 0 \\
  . & . & . & . & .. & . & . \\
  . & . & . & . & .. & . & . \\
  \bar{\kappa} & 0 & 0 & 0 & .. & \bar{\kappa} & -2\bar{\kappa}-k_0 \\
 \end{array}
 \right). \label{hc2}
 \ee
 The eigenvalues $\nu_i$ and eigenvectors $|\psi_i \rangle$ $(i=1,2 \dots N)$ of $M$ are obtained to be 
 \bea
 \nu_i &=& -k_0-4\bar{\kappa}\,\, \mathrm{sin}^2\left(\frac{\pi i}{N}\right) \cr
 \psi_{i}^{j} &=&  \frac{1}{\sqrt{N}}\left[\mathrm{cos}\left(\frac{2\pi i j}{N}\right)+\mathrm{sin}\left(\frac{2\pi i j}{N}\right)\right], \label{hc7}
 \eea
  with $\psi_{i}^{j}$ being the $j$-th component of $|\psi_i \rangle$. Consequently, the matrix in Eq. (\ref{hc2}) can be diagonalized as $M_d= U^{-1}MU$ where $(M_d)_{i,j}=\nu_i \delta_{i,j}$ and $U_{i,j}=\psi_{i}^{j}$. Also note, in this case $U^{-1}_{i,j}=U_{i,j}=\psi_{i}^{j}$. So, Eq. (\ref{hc1}) can now be expressed conveniently as 
  \be
 \frac{d^2 \delta \bold{Q}}{d\tau^2}=M_d \delta \bold{Q} -\bar{\gamma} \frac{d \delta \bold{Q}}{d\tau}, \label{hc3}
 \ee
 where $\delta \bold{Q} =\left(\delta q_1 \dots \delta q_i \dots \delta q_N\right)^{T}$ with $\delta q_i=\sum_{j}U^{-1}_{i,j} \delta y_i$. $\delta q_i$-s are uncoupled variables with individual equations of motions as below
 \be
  \frac{d^2 \delta q_i}{d\tau^2}+\bar{\gamma} \frac{d \delta q_i}{d\tau}-\nu_i \delta q_i =0. \label{hc4}
 \ee
 The above equation, being uncoupled, can be solved directly and resulting solution is given below
 \be
 \delta q_i(\tau) = \epsilon~ e^{\frac{-\bar{\gamma} \tau}{2}} U^{-1}_{i,\frac{N+1}{2}} \left[\mathrm{cos}(\Delta_i \tau)+\frac{\bar{\gamma}}{2\Delta_i}\mathrm{sin}(\Delta_i \tau)\right], \label{hc5}
 \ee
 where
 \be
 \Delta_j= \sqrt{4\bar{\kappa}~\mathrm{sin}^{2}\left(\frac{\pi j}{N}\right)+ k_0 -\left(\frac{\bar{\gamma}}{2}\right)^2}. \nonumber
 \ee
 We would be dealing with the under-damped scenario where $(\frac{\bar{\gamma}}{2})^2<k_0$. Now, we can invert $\delta q_i$ to obtain the following expression for $\delta y_i(\tau)=\sum_{j}U_{i,j} \delta q_i(\tau)$ as
 \be
\delta y_i(\tau)=\frac{\epsilon ~e^{-\frac{\bar{\gamma} \tau}{2}}}{N}\sum_{j=1}^{N}\psi_{i}^{j} \psi_{j}^{\frac{N+1}{2}} \left[\mathrm{cos}(\Delta_j \tau)+\frac{\bar{\gamma}}{2\Delta_j}\mathrm{sin}(\Delta_j \tau)\right].
 \label{hc6}
 \ee
 By using Eq. (\ref{hc7}) and making a shift in the oscillator index as $i\rightarrow(i-\frac{N+1}{2})$, we obtain the OTOC defined as $D(i,\tau)=\frac{\delta y_i}{\epsilon},$ is given by
 \bea
 D(i,\tau)=\frac{e^{-\frac{\bar{\gamma} \tau}{2}}}{N}\sum_{j=1}^{N}&&\left[\mathrm{cos}\left(\frac{2 \pi i j}{N}\right)+\mathrm{sin}\left(\frac{2 \pi j}{N}(i+1)\right)\right] \cr &\times& \left[\mathrm{cos}(\Delta_j \tau)+\frac{\bar{\gamma}}{2\Delta_j}\mathrm{sin}(\Delta_j \tau)\right],
 \label{hc_8}
 \eea
 for $-\frac{N-1}{2}\leqslant i \leqslant \frac{N-1}{2}$.  Now we note that 
\bea
&& \sum_{j=1}^{N} \mathrm{sin}\left(\frac{2 \pi j}{N}(i+1)\right) \left[\mathrm{cos}(\Delta_j \tau)+\frac{\bar{\gamma}}{2\Delta_j}\mathrm{sin}(\Delta_j \tau)\right] \cr &=& \sum_{j=1}^{N}\chi(i,j,\tau)=0, 
\label{hc_9}
\eea
using the fact $\chi(i,N-j,\tau)=-~\chi(i,j,\tau)$ for $j=1,2\dots (N-1)/2$ and $\chi(i,N,\tau)=0$. So, using (\ref{hc_9}), Eq. (\ref{hc_8}) reduces to
\bea
 && D(i,\tau)\cr &=& \frac{\mathrm{e}^{-\bar{\gamma} \tau/2}}{N}\sum_{j=1}^{N} \mathrm{cos}\left(\frac{2 \pi i j}{N}\right) \left[\mathrm{cos}(\Delta_j \tau)+\frac{\bar{\gamma}}{2\Delta_j}\mathrm{sin}(\Delta_j \tau)\right] \cr
 &=& \frac{\mathrm{e}^{-\frac{\bar{\gamma} \tau}{2}}}{N} \sum_{j=1}^{N} [\mathrm{cos}\left(\frac{2 \pi i j}{N}-\Delta_j \tau\right) \cr && ~~~~~~~ + \frac{\bar{\gamma}}{2\Delta_j}\mathrm{sin}\left(\frac{2 \pi i j}{N}-\Delta_j \tau\right)]. 
 \label{hc_10}
 \eea
 \begin{figure}[h]
  \centering
  \subfigure[]{\includegraphics[scale=0.50]{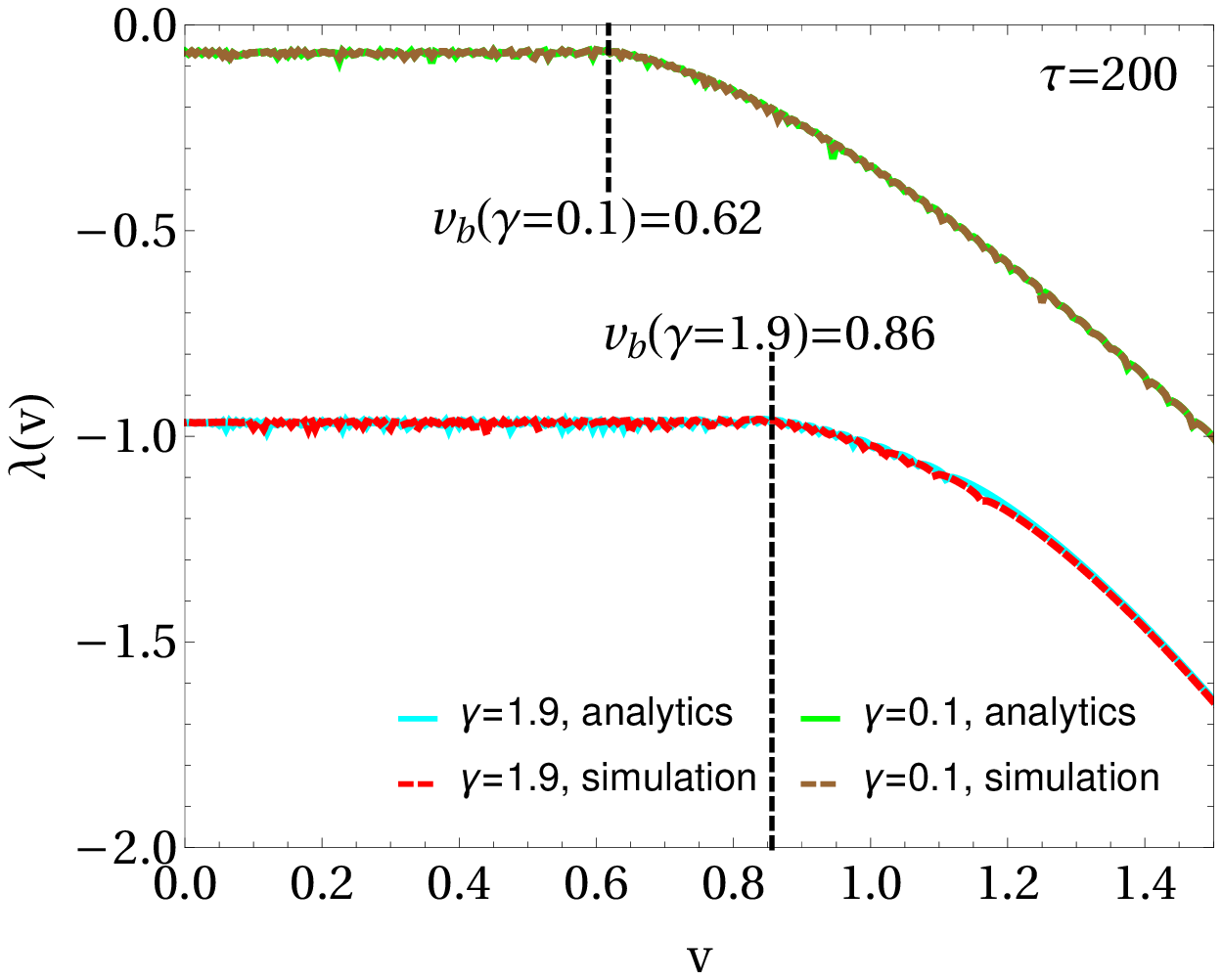}}\hfill
  \subfigure[]{\includegraphics[scale=0.50]{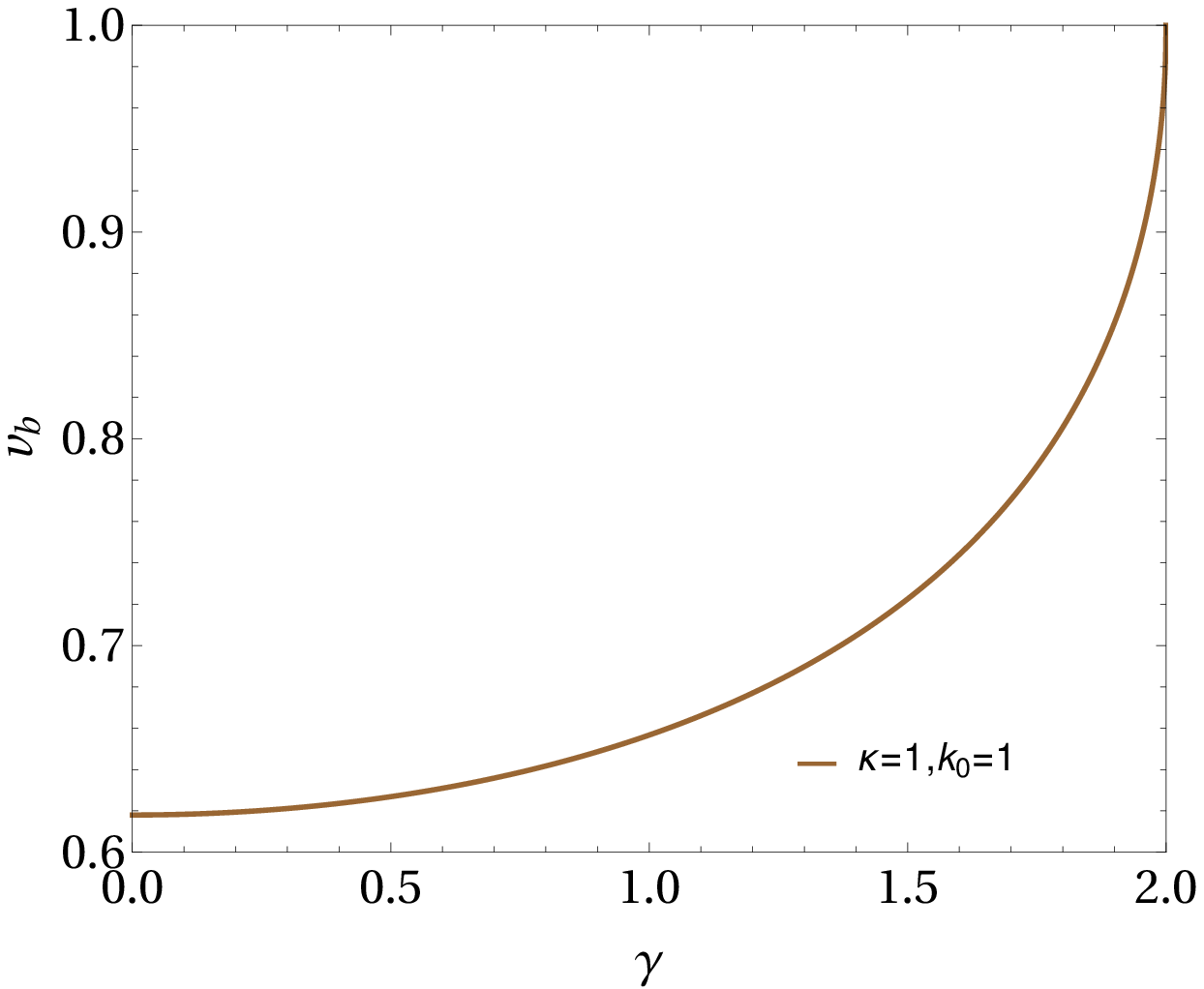}}
  \caption{In sub-figure (a), we have plotted VDLE vs  $v$ for two different dissipation values, $\gamma=0.1$ and $\gamma=1.9$, both from numerical computation (showed by the solid lines) and analytical calculation from Eq. (\ref{hc_10}) (showed by dotted lines) at time $\tau=200$. We observe that more the dissipation, more number of oscillators gain enough perturbation to reach the VDLE value $\lambda(v)\approx-\frac{\gamma}{2}$. This indicates that $v_b(\gamma)$ is larger for larger dissipation. This is indeed the case as clearly observed from sub-figure (b) where we have plotted $v_b$, from Eq. (\ref{hc_12}), as a function of $\gamma$.}
\label{fig:hcvbvsg}
\end{figure}
 Since we are considering a spatially extended system of very large size $N,$ in the limit $N\rightarrow\infty,$ we can take the continuum limit of Eq. (\ref{hc_10}) by identifying  $\frac{\pi j}{N}=q$ where $q\in [0,\pi]$ is a continuous variable. So, the sum in Eq. (\ref{hc_10}) becomes an integral as below
\bea
 && D(i=v\tau,\tau) = \frac{e^{-\frac{\bar{\gamma} \tau}{2}}}{\pi}  \int_{0}^{\pi} \mathrm{d}q \times \cr && \left[\mathrm{cos}\left(2\tau(qv-\frac{1}{2}\Delta_q)\right) -\frac{\bar{\gamma}}{2\Delta_q}\mathrm{sin}\left(2\tau(qv-\frac{1}{2}\Delta_q)\right)\right],~~~~~~~~
\label{hc_11}
\eea
with 
\bea
\Delta_q=\sqrt{2 \bar{\kappa}}\sqrt{1+\eta-\mathrm{cos}(2q)}
\eea
where $\eta=\frac{k_0-(\bar{\gamma}/2)^2}{2\bar{\kappa}}$. 
The integrand in the above equation (\ref{hc_11}) is like a forward moving wave with angular frequency $\omega(q)$ satisfying the dispersion relation $\omega(q)=\frac{1}{2}\Delta_q$. Then one can define the group velocity ($v_g$) and butterfly speed ($v_b$) from there as 
 \bea
 v_g(q) &=& \frac{\partial \omega}{\partial q}=\frac{\sqrt{\bar{\kappa}}~\mathrm{sin}(2q)}{\sqrt{2(1+\eta-\mathrm{cos}(2q)})} \cr
 v_b(\bar{\kappa},\bar{\gamma},k_0) &=& \mathrm{max}_{q}~ v_g \nonumber \\ &=& \sqrt{\bar{\kappa}}~\sqrt{1+\eta-\sqrt{(1+\eta)^2-1}}, 
 \label{hc_12}
\eea

where $\mathrm{max}_{q}~ v_g =v_g(q^\ast)$ with $q^\ast$ satisfying the equation $\mathrm{cos}(2q^\ast) = (1+\eta) - \sqrt{(1+\eta)^2-1}$. Clearly, at $q=q^\ast,$ $\frac{\partial^2 \omega}{\partial q^2}|_{q^\ast}=0$ implying that $q^\ast$ is the saddle point of $\omega(q)$ such that $\frac{\partial^3 \omega}{\partial q^3}|_{q^\ast}<0$.

Since we have the exact analytical expression of the OTOC for the DC in Eq. (\ref{hc_10}), we can directly calculate the VDLE using Eq. (\ref{vdle}). This is plotted for two different values of the dissipation in Fig.~\ref{fig:hcvbvsg}(a) at $\tau=200$, the corresponding data from simulation are presented in the same plot. The analytical and numerical data exhibit excellent match. Interestingly,  Fig.~\ref{fig:hcvbvsg}(a) reveals that more the dissipation value, more number of oscillators tend to achieve enough perturbation to attain $\lambda(v)\approx-\frac{\gamma}{2}$. This, in turn, indicates that the IS is larger for larger $\gamma$. This fact is further ensured by the plot of $v_b$ (calculated from Eq. (\ref{hc_12})) as a function of $\gamma$ presented in Fig.~\ref{fig:hcvbvsg}(b). There we clearly observe that $v_b$ is an increasing function of $\gamma$ for the driven dissipative harmonic chain. Although this might seem somewhat surprising, actually one have to keep in mind that in Fig.~\ref{fig:hcvbvsg}, what one measures is, how far a perturbation (however small it may be) can reach rather than the magnitude of the perturbation.

Note, in absence of dissipation ($\bar{\gamma}=0$) and on-site potential ($k_0=0$), we get, $\eta=0$ and the dispersion relation simplifies to $\omega(q)=\sqrt{\bar{\kappa}}~\mathrm{sin}(q)$. Consequently, for this conserved harmonically coupled chain,  the group velocity is $v_g=\sqrt{\bar{\kappa}}~\mathrm{cos}(q)$ and the butterfly speed simply becomes $v_b=\sqrt{\bar{\kappa}}$ occurring at $q=q^\ast=0$. 

Our goal is to analyze the behavior of $D(i,t)$ near $v\approx v_b$. To achieve that, we can do a saddle point approximation of the integral in (\ref{hc_11}) by analyzing the integrand near $q^\ast$, i.e., letting $q=q^\ast+\delta q$ where  $\delta q \in (-\epsilon,\epsilon)$, $\epsilon$ being a very small number. It is important to note that the previous statement is based on the underlying assumption that $q^\ast \in (0,\pi)$. In other words, the endpoints $q^\ast=0$ and $q^\ast=\pi$ have to be dealt with separately since for them the  neighborhoods  are restricted only to $\delta q \in (0,\epsilon)$ and $\delta q \in (-\epsilon, 0)$ respectively. In context of system parameters, the equation $\mathrm{cos}(2q^\ast) = (1+\eta) - \sqrt{(1+\eta)^2-1}$ directly implies that $q^\ast=0$ and $q^\ast=\pi$ means $\eta=0$. An example system leading to this scenario is $k_0=0=\gamma$, i.e., the chain of harmonically coupled oscillators in absence of dissipation and on-site harmonic potential. The analysis for this case will be done separately at the end of this section.  For now, we stick to the general driven dissipative coupled harmonic chain for which $\eta\neq 0$. Near  $v\approx v_b$, from Eq. (\ref{hc_11}), the OTOC becomes
\bea
 D(v\tau\approx v_b \tau,\tau)&=&\frac{e^{-\frac{\gamma \tau}{2}}}{\pi} [g(q^\ast)\int_{-\epsilon}^{\epsilon} \mathrm{d}(\delta q)~ \mathrm{cos}[2\tau~ h\delta q)] \cr &+& \bar{g}(q^\ast)\int_{-\epsilon}^{\epsilon} \mathrm{d}(\delta q)~ \mathrm{sin}[2\tau~ h(\delta q)]], 
 \label{hc_13}
\eea
where
\bea
 h(\delta q)&=&(v-v_b)\delta q +|\frac{1}{2}\frac{\partial^3 \omega}{\partial q^3}|\frac{(\delta q)^3}{3} \cr &=&(v-v_b)\delta q +2 v_b \frac{(\delta q)^3}{3}~, \cr
  g(q^\ast,\tau)&=&\mathrm{cos} \big(2\tau(vq^\ast-\frac{1}{2}\Delta_{q^\ast})\big)\cr&-&\frac{\bar{\gamma}}{2\Delta_{q^\ast}} \mathrm{sin}\big(2\tau(vq^\ast-\frac{1}{2}\Delta_{q^\ast})\big)~,\cr
   \bar{g}(q^\ast,\tau)&=&-\mathrm{sin}\big(2\tau(vq^\ast-\frac{1}{2}\Delta_{q^\ast})\big)\cr&-&\frac{\bar{\gamma}}{2\Delta_{q^\ast}} \mathrm{cos}\big(2\tau(vq^\ast-\frac{1}{2}\Delta_{q^\ast})\big). 
 \label{hc_44}
\eea
\begin{figure}[h]
  \centering
  \subfigure[]{\includegraphics[scale=0.34]{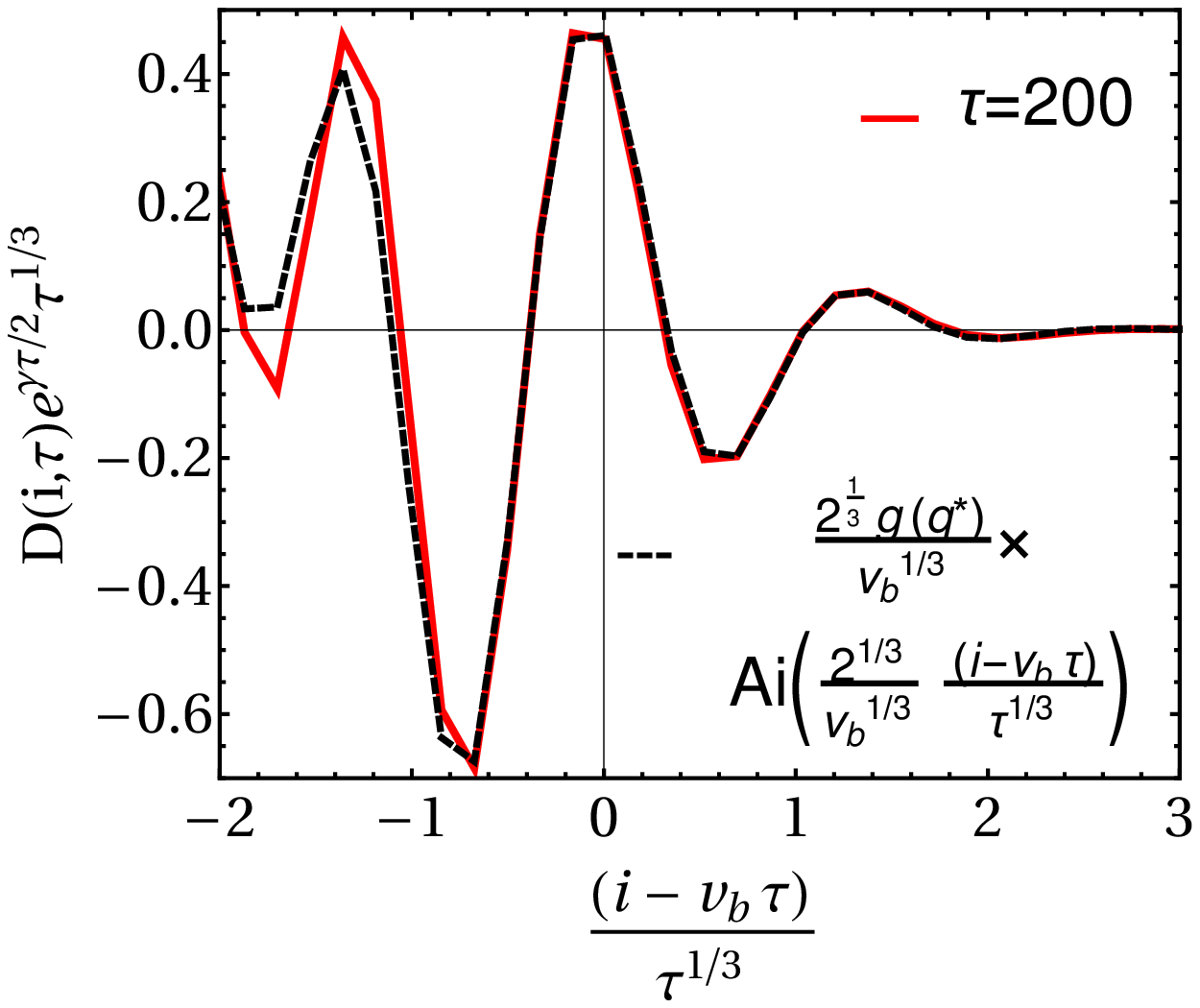}}\hfill
  \subfigure[]{\includegraphics[scale=0.34]{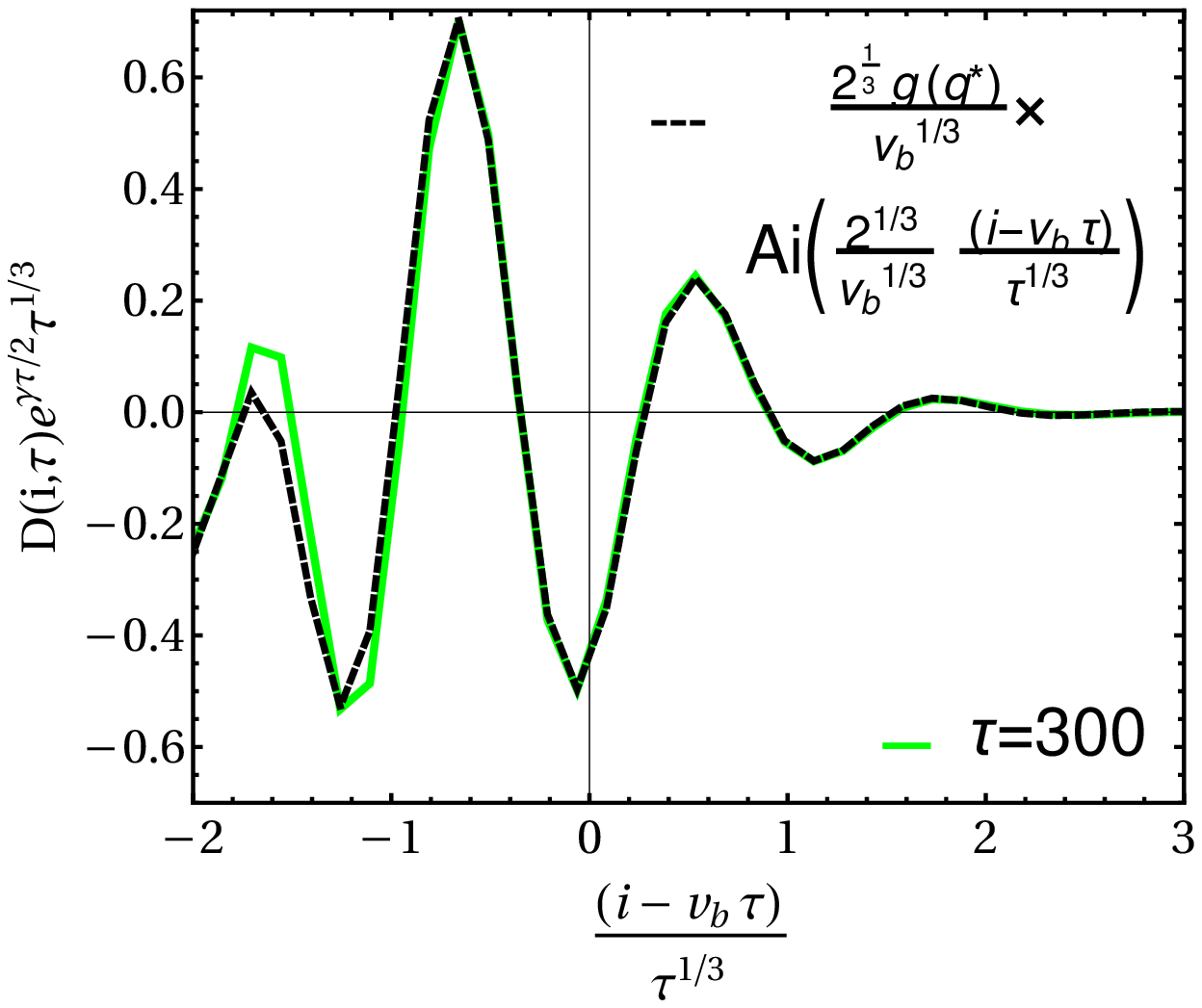}}\\
  \subfigure[]{\includegraphics[scale=0.34]{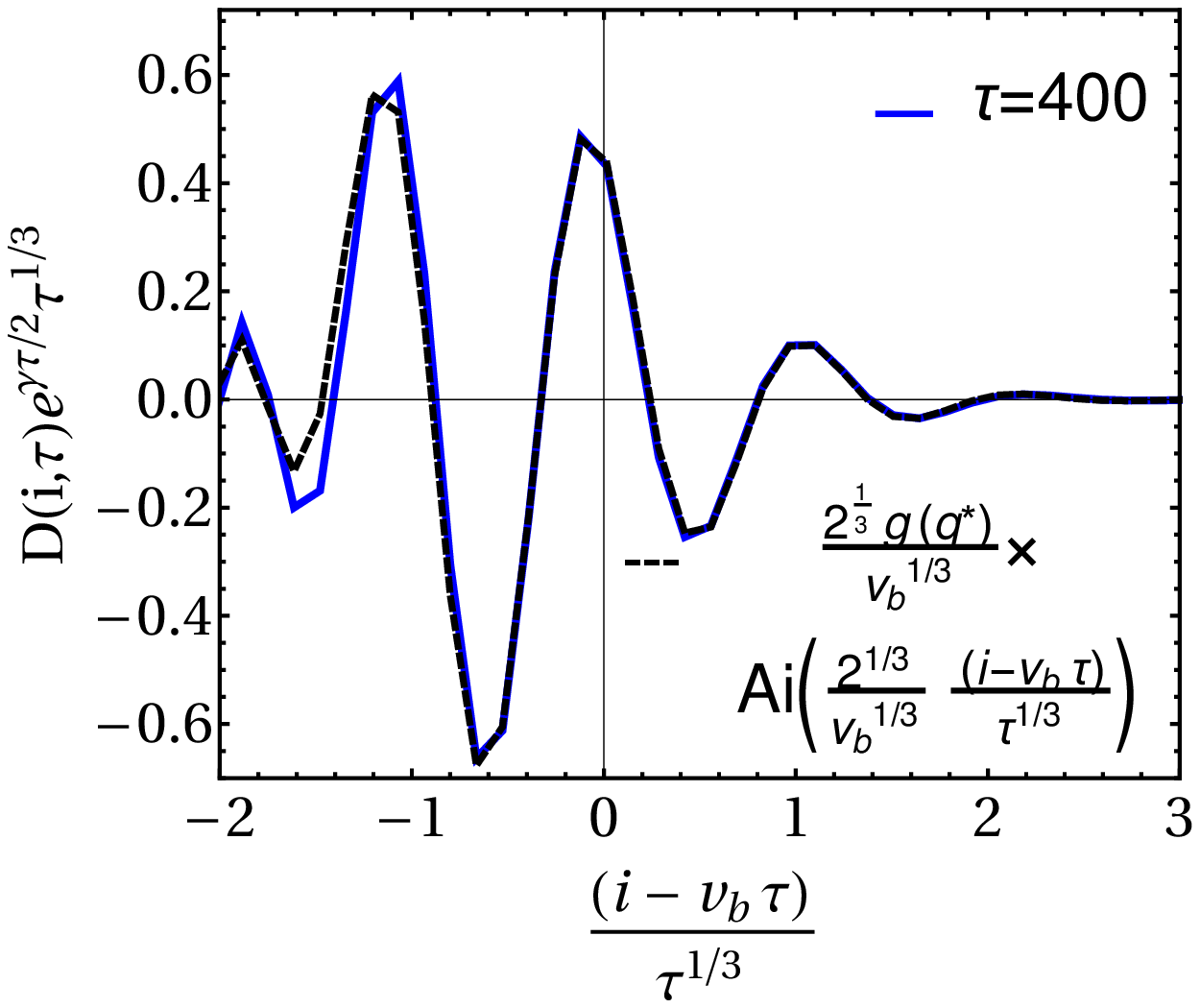}}\hfill
  \subfigure[]{\includegraphics[scale=0.34]{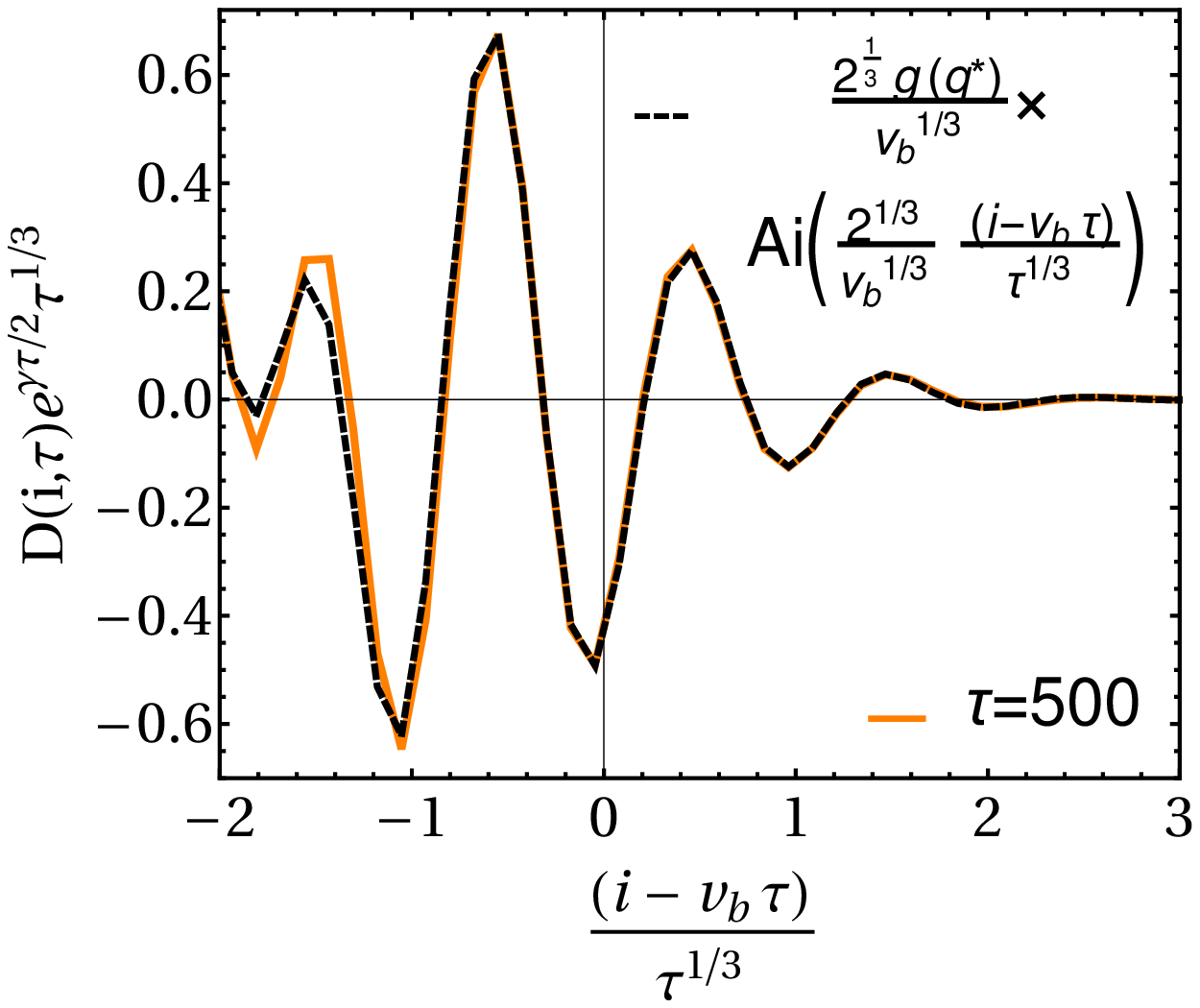}}
  \caption{$\left\lbrace \kappa=1, k_0=0.1, \gamma=0.05 \right\rbrace$. In the sub-figures of the above figure, we have plotted the exact form of OTOC (scaled by $e^{\gamma\tau/2}\tau^{1/3}$) versus the scaled variable $(i-v_b \tau)/\tau^{1/3}$ from Eq. (\ref{hc_10}) for a driven dissipative coupled harmonic chain at different $\tau$. We compare these with the corresponding approximate expression of OTOC obtained in Eq. (\ref{hc17}) using continuum theory. The observation is that at each $\tau,$ the exact [shown by solid curve computed from Eq. (\ref{hc_10})] and approximate [shown by dashed curve computed from Eq. (\ref{hc_10})] expressions of OTOC show perfect agreement near $v\approx v_b$ and start deviating from each other as we go reasonably far from $v\approx v_b$.}
\label{fig:hcg}
\end{figure}

Now using the fact that $\int_{-a}^{a}\mathrm{d} x f(x)= \int_{0}^{a}\mathrm{d} x [f(x)+f(-x)]$ and $\mathrm{sin}[2\tau~ h(\delta q)]$ is an odd function, the integral in Eq. (\ref{hc_13}) reduces to 
\bea
  &&
  D(i=v\tau\approx v_b \tau,\tau) \cr \nonumber \\ 
  &=&\frac{2 e^{-\frac{\bar{\gamma} \tau}{2}}}{\pi} g(q^\ast,\tau)\int_{0}^{\epsilon} \mathrm{d}(\delta q)~ \mathrm{cos}[2t~ h(\delta q)] \cr
  &=& \frac{2 e^{-\frac{\bar{\gamma} \tau}{2}} g(q^\ast,\tau)}{\pi} \int_{0}^{\epsilon} \mathrm{d}(\delta q) ~\mathrm{cos} \left[2(v-v_b)\tau~\delta q + 4 v_b \tau \frac{(\delta q)^3}{3}\right]. \nonumber
\eea
Now, with the following variable transformation
\be
 (4 v_b \tau)^{1/3}~\delta q=s ~~~~~ \Rightarrow s \in (0,\infty) ~~\mathrm{as}~ \tau\rightarrow\infty, \nonumber
\ee
the above integral  becomes
 \bea
  && D(v\tau\approx v_b \tau,\tau) = \frac{2 e^{-\frac{\bar{\gamma} \tau}{2}} g(q^\ast,\tau)}{(4 v_b \tau)^{1/3} \pi} \times \cr && \int_{0}^{\infty} \mathrm{d}s ~~\mathrm{cos} \left[\frac{s^3}{3}+ \frac{2^{\frac{1}{3}}(v-v_b)\tau^{\frac{2}{3}}}{v_b^{\frac{1}{3}}}  s \right].
 \label{hc_16}
 \eea
The above integral is in the form of the well-known Airy integral. So, we finally have
\begin{eqnarray}
 D(v\tau,\tau)&=& 
 \begin{cases}
& \frac{2 e^{-\frac{\bar{\gamma} \tau}{2}} g(q^\ast,\tau)}{(4v_b \tau)^{1/3}} \mathrm{Ai}(z),~~~~\text{for}~v \gtrsim v_b \\
&\frac{2 e^{-\frac{\bar{\gamma} \tau}{2}} g(q^\ast,\tau)}{(4v_b \tau)^{1/3}} \mathrm{Ai}(-z),~~\text{for}~v \lesssim v_b
\end{cases}
\label{hc17} \\
\text{with} && z=\frac{2^{\frac{1}{3}}|v_b-v|\tau^{\frac{2}{3}}}{v_b^{\frac{1}{3}}} >0. \label{hc35}
\end{eqnarray}
\begin{figure}[h]
  \centering \includegraphics[width=8 cm]{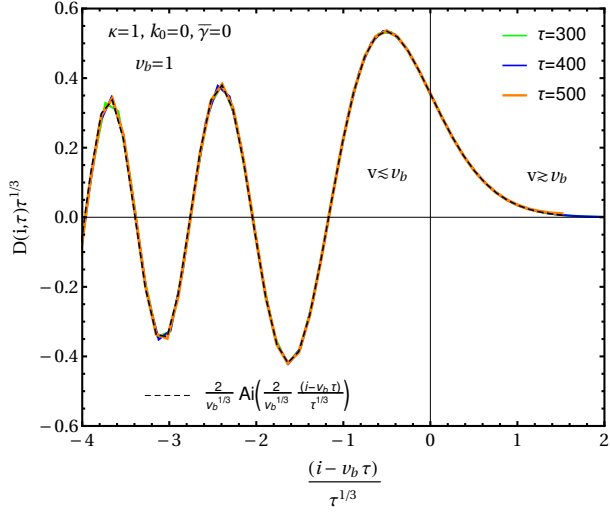}
  \caption{$\left\lbrace \kappa=1, k_0=0, \gamma=0 \right\rbrace$. Here we compare the exact expression of the OTOC in Eq. (\ref{hc_10}) for a harmonically coupled Duffing chain (in absence of dissipation and on-site harmonic potential) to the approximate expression of OTOC in Eq. (\ref{hc17}) near $v\approx v_b$ obtained using continuum  theory. The OTOC (scaled by a factor $\tau^{1/3}$) at different times [shown by the solid curves computed from Eq. (\ref{hc_10})] collapse perfectly on top of each other when plotted against the scaled variable $(i-v_b \tau)/\tau^{1/3}$. This data collapse also exhibits perfect agreement with the corresponding continuum approximation [the dashed curve computed from  Eq. (\ref{hc17})] within a reasonably large range about $v \approx v_b$. }
\label{fig:hcg0}
\end{figure}
In order to compare the exact expression of the OTOC in Eq. (\ref{hc_10}) with the corresponding approximated expression near $v\approx v_b$ in Eq. (\ref{hc17}), we plot $D(i,\tau)$ (scaled by $\tau^{\frac{1}{3}}$) as a function of $(i-v_b \tau)/\tau^{1/3}$ in Fig.~\ref{fig:hcg} for an arbitrary chosen parameter set $\left\lbrace \kappa=1, k_0=0.1, \gamma=0.05 \right\rbrace$. The four panels in  Fig.~\ref{fig:hcg} correspond to different $\tau$ (sufficiently large). We observe that, for each $\tau$, the exact form of OTOC (shown by solid curve) computed from Eq. (\ref{hc_10}) exhibits prefect agreement near $v\approx v_b$ with the corresponding approximate form (shown by dashed curve), obtained using continuum approximation,  in Eq. (\ref{hc17}). However, as one moves sufficiently far from $v\approx v_b$ region, the two expressions [Eq. (\ref{hc_10}) and Eq. (\ref{hc17})] start deviating from one other.

Since we are in the limit $\tau\rightarrow\infty$, from Eq. (\ref{hc17}), we can use the large $z$ asymptotic of Airy functions,  we have
\begin{align}
D(v\tau,\tau)=
\begin{cases}
\frac{ \hat{g}(q^\ast,\tau)}{2\sqrt{\tau}}~~ e^{-\frac{\bar{\gamma}}{2}\tau-\frac{2^{\frac{5}{2}}}{3\sqrt{v_b}}\tau(v-v_b)^{\frac{3}{2}}},~~~~~~~~~~~~v>v_b \\
\frac{e^{-\frac{\bar{\gamma} \tau}{2}} \hat{g}(q^\ast,\tau)}{\sqrt{\tau}}~ \mathrm{sin}\left[\frac{\pi}{4}+\frac{2^{\frac{5}{2}}\tau}{3\sqrt{v_b}}(v-v_b)^{\frac{3}{2}}\right],v<v_b 
\end{cases}
\label{hc_50}
\end{align}
where $\hat{g}(q^\ast,\tau)=\frac{g(q^\ast)}{2^{-\frac{3}{4}}\sqrt{\pi}(v-v_b)^{\frac{1}{4}}v_b^{\frac{1}{4}}}$. It should be mentioned that, in Eq. (\ref{hc_50}), both $\hat{g}(q^\ast,\tau)$ and $v_b$ are functions of $\bar{\gamma}$ through $q^\ast(\bar{\gamma})$ and $\eta(\bar{\gamma})$ respectively. So, other than the explicit exponential dependence as $e^{-\frac{\bar{\gamma}}{2}\tau},$ the OTOC depends non-trivially on the dissipation through $\hat{g}(q^\ast,\tau)$ and $v_b$.  
The velocity dependent Lyapunov exponent (VDLE), $\lambda(v)$ is defined in Eq. (\ref{vdle}) in the main text. Using this, we obtain that near $v\approx v_b$, the VDLE are given by
\bea
\lambda(v) -\lambda &\approx& -(v-v_b)^{\frac{3}{2}} ~~~\mathrm{for}~~ v>v_b \cr
\lambda(v) &\approx& \lambda  ~~~ \mathrm{for}~~ v<v_b~, 
\eea
where $\lambda=-\frac{\gamma}{2}$.

As stated earlier, we would now like to consider the special case of harmonically coupled chain without any dissipation and on-site potential (i.e. $k_0=0=\gamma$). For this chain, we have $\eta=\frac{k_0-(\bar{\gamma}/2)^2}{2\bar{\kappa}}=0$ leading to $q^\ast=0$ or $q^\ast=\pi$. Without any loss of generality we consider $q^\ast=0$ so that now $\delta q \in (0,\epsilon)$ instead of $\delta q \in (-\epsilon,\epsilon)$. Correspondingly, after a saddle point approximation, Eq. (\ref{hc_12}) in this case boils down to
\bea
 \hspace*{-0.5 cm} D(i=v\tau\approx v_b \tau,\tau)&=&\frac{1}{\pi}\int_{0}^{\epsilon} \mathrm{d}(\delta q)~ \mathrm{cos}[2\tau~ h(\delta q)] 
 \label{hc_31}
\eea
where
\bea
 h(\delta q)=(v-v_b)\delta q +v_b \frac{(\delta q)^3}{6} .
 \label{hc_32}
\eea
Note that, for a harmonically coupled chain, $ g(q^\ast,\tau)=1$ and $ \bar{g}(q^\ast,\tau)=0$. In the limit $\tau\rightarrow \infty,$ with the variable transformation $(v_b \tau)^{1/3} \delta q=s,$ Eq. (\ref{hc_31}) transforms into
\bea
  && D(v\tau\approx v_b \tau,\tau) = \frac{2}{(v_b \tau)^{1/3} \pi} \times \cr && \int_{0}^{\infty} \mathrm{d}s ~~\mathrm{cos} \left[\frac{s^3}{3}+ \frac{2(v-v_b)\tau^{\frac{2}{3}}}{v_b^{\frac{1}{3}}}  s \right].
 \label{hc_33}
 \eea
 As already discussed, the integral in Eq. (\ref{hc_33}) is in the well-known form of Airy Function, we have 
 \begin{eqnarray}
 D(v\tau,\tau)&=& 
 \begin{cases}
& \frac{2}{(v_b \tau)^{1/3}} \mathrm{Ai}(z),~~~~\text{for}~v \gtrsim v_b \\
&\frac{2}{(v_b \tau)^{1/3}} \mathrm{Ai}(-z),~~\text{for}~v \lesssim v_b 
\end{cases}
\label{hc34} \\
\text{with} && z=\frac{2|v_b-v|\tau^{\frac{2}{3}}}{v_b^{\frac{1}{3}}} >0. \label{hc36}
\end{eqnarray}
Note the difference between the expressions of OTOC for the driven dissipative coupled harmonic chain in Eq. (\ref{hc17}), (\ref{hc35}) with that of a harmonically coupled chain in Eq. (\ref{hc34}), (\ref{hc36}). Due to the absence of the time dependent term $g(q^\ast,\tau)$ in Eq. (\ref{hc34}), we expect collapse of data at different $\tau$ when the OTOC $D(i,\tau)$ is scaled by $\tau^{\frac{1}{3}}$. This is indeed observed in Fig.~\ref{fig:hcg0} where the exact expression for OTOC (scaled by $\tau^{\frac{1}{3}}$) in Eq. (\ref{hc_10}) for a harmonically coupled chain (with $\kappa=1$) is plotted against the scaled variable $(i-v_b \tau)/\tau^{1/3}$. Apart from the excellent data collapse at different $\tau$ (shown by the solid curves), Fig.~\ref{fig:hcg0} exhibits perfect agreement with the exact OTOC expression in Eq. (\ref{hc_10}) to the corresponding approximated expression (obtained through continuum theory and saddle point approximation) in a considerably large range around $v\approx v_b$ in Eq. (\ref{hc17}) (shown by the dashed curve).


\begin{thebibliography}{99}
\bibitem{Lorenz_1993} E. Lorenz, {\it The Essence of CHAOS},  1993 University of Washington Press, Seattle, Washington.

\bibitem{Strogatz_1994} S. H. Strogatz, {\it Nonlinear Dynamics and Chaos}, 1994 Perseus Books Publishing, Reading, Massachusetts.

\bibitem{Gutzwiller_1990} M. C. Gutzwiller, {\it Chaos in Classical and Quantum Mechanics}, 1990 Springer-Verlag, New York.

\bibitem{Haake_1991} F. Haake, {\it Quantum Signatures of Chaos}, 1991 Spring-Verlag, Berlin.

\bibitem{Zeng_1993} X. Zeng, R. A. Pielke and R. Eykholt, Bulletin of the American Meteorological Society {\bf 74}(4), 631 (1993).  

\bibitem{Selvam_2010} A. M. Selvam, arXiv:1006.4554 (2010).

\bibitem{Huppert_1998} A. Huppert and L. Stone, The American Naturalist {\bf 152}(3), 447 (1998).

\bibitem{Field_1993} {\it Chaos in Chemistry and Biochemistry}, ed. by R. J. Field and L. Gy\"{o}rgyi, 1993 World Scientific Publishing Co. Pte. Ltd., Singapore.

\bibitem{Gaspard_1999} P. Gaspard, Physica A {\bf 263}, 315 (1999).

\bibitem{Gaspard_1998} P. Gaspard, M.E. Briggs, M. K. Francis,J V. Sengers, R. W. Gammon, J. R. Dorfman and R. V. Calabrese, Nature {\bf 394}, 865 (1998).

\bibitem{Srivastava_2013} R. Srivastava, P. K. Srivastava and J. Chattopadhyay, Eur. Phys. J. Special Topics {\bf 222}, 777 (2013). 

\bibitem{Skinner_1994} J. E. Skinner, Nature Bio/Technology {\bf 12}, 596 (1994).

\bibitem{Ditto_1996} W. L. Ditto, AIP Conference Proceedings {\bf 376}, 175 (1996). 

\bibitem{Lesne_2006} A. Lesne, Riv Biol. {\bf 99}(3), 467 (2006).

\bibitem{Lloyd_1995} A. L. Lloyd and D. Lloyd, Biological Rhythm Research {\bf 26}(2), 233 (1995). 

\bibitem{Tchana_2008} A. S. Tchana, P. Woafo and R. Yamapi, International Journal of Bifurcation and Chaos {\bf 18}(11), 3473 (2008).

\bibitem{Kwuimy_2008} C. A. K. Kwuimy and P. Woafo, Nonlinear Dyn. {\bf 53}, 201 (2008).

\bibitem{Bao_2011} B. Bao, Z. Ma, J. Xu, Z. Liu and Q. Xu, International Journal of Bifurcation and Chaos {\bf 21}(9), 2629 (2011).

\bibitem{Chau_2011} K. T. Chau and Z. Wang,{\it Chaos in Electric Drive Systems}, 2011 John Wiley \& Sons (Asia) Pte Ltd, Asia.

\bibitem{Wei_1997} J. G. Wei and G. Leng, Applied Mathematics and Computations {\bf 88}, 77 (1997).

\bibitem{Kozlowski_1995} J. Kozlowski, U. Parlitz and W. Lauterborn, Phys. Rev. E {\bf 51}(3), 1861 (1995).

\bibitem{Vastano_1988} J. A. Vastano and H. L. Swinney, Phys. Rev. Lett. {\bf 60},  1773 (1988).

\bibitem{Wacker_1995} A. Wacker, S. Bose and E. Sch\"oll, Europhys. Lett. {\bf 31}, 257 (1995).

\bibitem{Lepri_1996} S. Lepri, A. Politi and A. Torcini, J. Stat. Phys. {\bf 82}, 1429 (1996).

\bibitem{Lepri_1997} S. Lepri, A. Politi and A. Torcini, J. Stat. Phys. {\bf 88}, 31 (1997).

\bibitem{Giacomelli_2000} G. Giacomelli, R. Hegger, A. Politi and M. Vassalli, Phys. Rev. Lett. {\bf 85}, 3616 (2000).

\bibitem{Pazo_2016} D. Pazo, J. M. Lopez and A. Politi, Phys. Rev. Lett. {\bf 117}, 034101 (2016).

\bibitem{Stahlke_2011} D. Stahlke and R. Wackerbauer, Phys. Rev. E {\bf 83}, 046204 (2011).

\bibitem{Chopra_2009} N. Chopra and M. W. Spong, IEEE Transactions on Automatic control {\bf 54}, 353 (2009).

\bibitem{Mohanty_2004} P. K. Mohanty, Phys. Rev. E {\bf 70}, 045202(R) (2004).

\bibitem{Mohanty_2006} P. K. Mohanty and A. Politi, J. Phys. A : Math. Gen. {\bf 39}, L415 (2006).

\bibitem{Das_2018} A. Das, S. Chakrabarty, A. Dhar, A. Kundu, D. A. Huse and R. Moessner, Phys. Rev. Lett. {\bf 121}, 024101 (2018).

\bibitem{Khemani_2018} V. Khemani, D. A. Huse and A. Nahum, Phys. Rev. B. {\bf 98}, 144304 (2018).

\bibitem{Kumar_2019} D. Kumar, S. Bhattacharjee and S. S. Ray, arXiv:1906.00016 (2019). 

\bibitem{Bilitewski_2018} T. Bilitewski, S. Bhattacharjee and R. Moessner, Phys. Rev. Lett. {\bf 121}, 250602 (2018).

\bibitem{Sekino_2008} Y. Sekino and L. Susskind, J. High Energy Phys. {\bf 10}, 065 (2008).

\bibitem{Shenker_2014} S. H. Shenker and D. Stanford, J. High Energy Phys. {\bf 03}, 067 (2014).

\bibitem{Rozenbaum_2017} E. B. Rozenbaum, S. Ganeshan, and V. Galitski, Phys. Rev. Lett. {\bf 118}, 086801 (2017).

\bibitem{Kukuljan_2017} I. Kukuljan, S. Grozdanov, and T. Prosen, Phys. Rev. B {\bf 96}, 060301 (2017).

\bibitem{Bohrdt_2017} A. Bohrdt, C. Mendl, M. Endres, and M. Knap, New J. Phys. {\bf 19}, 063001 (2017).

\bibitem{Lakshminarayan_2019} A. Lakshminarayan, Phys. Rev. E {\bf 99}, 012201 (2019).

\bibitem{Zhang_2019} Y-L Zhang, Y. Huang and X. Chen, Phys. Rev. B {\bf 99}, 014303 (2019).

\bibitem{Loga_2019} B. Chakrabarty, S. Chaudhuri and R. Loganayagam, J. High Energy Phys. {\bf 2019}, 102 (2019).

\bibitem{He_2017} R-Q He and Z-Y Lu, Phys. Rev. B {\bf 95}, 054201 (2017).

\bibitem{Fan_2017} R. Fan, P. Zhang, H. Shen and H. Zhai, Science Bulletin {\bf 62}, 707 (2017).

\bibitem{Ray_2018} S. Ray, S. Sinha and K. Sengupta, Phys. Rev. A {\bf 98}, 053631 (2018).

\bibitem{Polchinski_2016} J. Polchinski and V. Rosenhaus, J. High Energy Phys. {\bf 2016}, 1 (2016).

\bibitem{Maldacena_2016} J. Maldacena and D. Stanford, Phys. Rev. D {\bf 94}, 106002 (2016).

\bibitem{Lin_2018} C-J Lin and O. J. Motrunich, Phys. Rev. B {\bf 97}, 144304 (2018).

\bibitem{Gopalakrishnan_2018} S. Gopalakrishnan, Phys. Rev. B {\bf 98}, 060302(R) (2018).

\bibitem{McGinley_2019} M. McGinley, A. Nunnenkamp and J. Knolle, Phys. Rev. Lett. {\bf 122}, 020603 (2019).

\bibitem{Chen_2019} X. Chen and T. Zhou, Phys. Rev. B. {\bf 100}, 064305 (2019). 

\bibitem{Duffing_1918} G. Duffing, {\it Erzwungene Schwingungen bei ver\"{a}nderlicher Eigenfrequenz und ihretechnische Bedeutung}, 1918.

\bibitem{Ueda_1978} Y. Ueda, Trans. IEE Japan {\bf 98-A}, 167 (1978) (in Japanese); English translation, Int. Jour. Non-linear Mech. {\bf 20}, 481 (1985).

\bibitem{Ueda_1979} Y. Ueda, J. Stat. Phys. {\bf 20}(2), 181 (1979).

\bibitem{Ueda_1991} Y. Ueda, Chaos, Solitons \& Fractals {\bf 1}(3), 199 (1991).

\bibitem{Stupnicka_1987} W. Szemplinska-Stupnicka, Journal of Sound and Vibration {\bf 113}(1), 155 (1987).

\bibitem{Englisch_1991} V. Englisch and W. Lauterborn, Phys. Rev. A {\bf 44}(2), 916 (1991).

\bibitem{Kovacic_2011} {\it The Duffing Equation}, ed. I. Kovacic and M. J. Brennan, 2011 John Wiley \& Sons Ltd., United Kingdom.

\bibitem{Gottwald_1992} J. A. Gottwald, L. N. Virgin and E. H. Dowell, Journal of Sound and Vibration {\bf 158}(3), 447 (1992).

\bibitem{Chabreyrie_2011} R. Chabreyrie and N. Aubry, arXiv:1108.4118 (2011).

\bibitem{Kenfack_2003} A. Kenfack, Chaos, Solitons and Fractals {\bf 15}, 205 (2003).

\bibitem{Musielak_2005} D. E. Musielak, Z. E. Musielak and J. W. Benner, Chaos , Solitons and Fractals {\bf 24}, 907 (2005).

\bibitem{Jothimurugan_2016} R. Jothimurugan, K. Thamilmaran, S. Rajasekar and M. A. F. Sanjuan, Nonlinear Dynamics {\bf 83}(4), 1803 (2016).

\bibitem{Wei_2011} X. Wei, M. F. Randrianadrasana, M. Ward and D. Lowe, Mathematical problems in Engineering {\bf 2011}, 1-16 (2011).

\bibitem{Kapitaniak_1993} T. Kapitaniak, Phys. Rev. E. {\bf 47}(5), R2975 (1993).

\bibitem{Lai_1994} Y-C. Lai and R. L. Winslow, Physica D {\bf 74}, 353 (1994).

\bibitem{Clerc_2018} M. G. Clerc, S. Coulibaly, M. A. Ferr\'e and R. G. Rojas, Chaos {\bf 28}, 083126 (2018).

\bibitem{Zapateiro_2013} M. Zapateiro, Y. Vidal, and L. Acho, IFAC Proc. Vol. {\bf 9}(1), 749 (2013).

\bibitem{Murali_1993} K. Murali and M. Lakshmanan, Phys. Rev. E {\bf 48}(3), R1624 (1993).

\bibitem{Hu_2003} N. Q. Hu and X. S. Wen, Journal of Sound and Vibration {\bf 268}, 917 (2003).

\bibitem{Zhang_2017} Y. Zhang, H. Mao, H. Mao and Z. Huang, Results in Physics {\bf 7}, 3243 (2017).

\bibitem{Liu_2011} X. Liu and X. Liu, Journal of Computers {\bf 6}(2), 359 (2011).

\bibitem{Umberger_1989} D. K. Umberger, C. Grebogi, E. Ott and B. Afeyan, Phys. Rev. A {\bf 39}(9), 4835 (1989).

\bibitem{Lieb_1972} E. H. Lieb and D. W. Robinson, Commun. Math. Phys. {\bf 28}, 251 (1972).

\bibitem{Deissler_1984} R. J. Deissler, Phys. Lett. A {\bf 100}, 451 (1984).

\bibitem{Kaneko_1986} K. Kaneko, Physica D: Nonlinear Phenomena {\bf 23}, 436 (1986).

\bibitem{Deissler_1987} R. J. Deissler and K. Kaneko, Phys. Lett. A {\bf 119}, 397 (1987).

\bibitem{Peng_1996} J. H. Peng, E. J. Ding, M. Ding and W. Yang, Phys. Rev. Lett. {\bf 76}, 904 (1996).

\bibitem{Lakshmanan_1996} M. Lakshmanan and K. Murali, {\it Chaos in Nonlinear Oscillators}. ed. L. O. Chua, 1996 World Scientific Publishing Co Pte Ltd, Singapore.

\bibitem{Blakely_2000} J. N. Blakely and D. J. Gauthier, International Journal of Bifurcation and Chaos {\bf 10}(4), 835 (2000).

\bibitem{Ivancevic_2007} V. G. Ivancevic and T. T. Ivancevic, {\it High Dimensional Chaotic and Attractor Systems}, 2007 Dordrecht, Springer-Verlag.

\bibitem{Musielak_2009} Z. E. Musielak and D. E. Musielak, International Journal of Bifurcation and Chaos {\bf 19}(9), 2823 (2009).

\bibitem{Kurchan_2016} J. Kurchan, J. Stat.Phys {\bf 171}, 965 (2018).

\bibitem{Roy_2001} A. Roy and J. K. Bhattacharjee, Phys. Lett. A {\bf 288}, 1 (2001).

\bibitem{Pokharel_2018} B. Pokharel, M. Z. R. Misplon, W. Lynn, P. Duggins, K. Hallman, D. Anderson, A. Kapulkin and A. K. Pattanayak, Sci Rep {\bf 8}, 2108 (2018).
\end{thebibliography}
\end{document}